\documentclass[11pt,onecolumn]{IEEEtran}

\pdfoutput=1
\usepackage[final]{pdfpages}
\usepackage{amsfonts,amssymb,amsmath}
\usepackage{graphicx}
\usepackage{epsfig,array,cite}
\usepackage{epsf,exscale,times}
\usepackage{color,multicol}

\let \bd = \textbf

\let \it = \textit
\let \t  = \text

\def \p1#1{#1^{-1}}

\def \~#1{\tilde{#1}}

\let \nn = \nonumber

\begin{document}
\title{Sparse Transmit Array Design for Dual-Function Radar Communications by Antenna Selection}
\author{Xiangrong~Wang, Aboulnasr~Hassanien, and Moeness~G.~Amin
\thanks{Xiangrong Wang is with School of Electronic and Information Engineering, Beihang University, Beijing, China, 100191. E-mail: xrwang@buaa.edu.cn.
Aboulnasr Hassanien is with the Department of Electrical Engineering, Wright State University, Dayton, OH 45435, USA. E-mail: hassanien@ieee.org. Moeness Amin is with Center for Advanced Communications, Villanova University, PA, USA, 19085. E-mail: moeness.amin@villanova.edu.
}
\thanks{The work by X Wang is supported by National Natural Science Foundation of China under Grant No. 61701016.}
}
\maketitle

\begin{abstract}

Dual-function radar communications (DFRC) systems have recently been proposed to enable the coexistence of radar and wireless communications, which in turn alleviates the increased spectrum congestion crisis. In this paper, we consider the problem of sparse transmit array design for DFRC systems by antenna selection where same or different antennas are assigned to different functions. We consider three different types of DFRC systems which implement different simultaneous beamformers associated with single and different sparse arrays with shared aperture. We utilize the array configuration as an additional spatial degree of freedom (DoF) to suppress the cross-interference and facilitate the cohabitation of the two system functions. It is shown that the use of sparse arrays adds to improved angular resolution with well-controlled sidelobes on DFRC system paradigm. The utilization of sparse arrays in DFRC systems is validated using simulation examples. 

\end{abstract}


\IEEEpeerreviewmaketitle

\section{Introduction}

In recent years, radio frequency (RF) spectrum is becoming increasingly congested with an exponentially growing demand by end-consumers. Consequently, defense applications are losing spectrum to commercial communications, and have to operate in contested environments. Emerging research in multi-function platforms aims at using common or shared aperture and frequency spectrum between radar, electronic warfare, and military communications \cite{Griffiths2013, Deng2013, Bliss2014, Hayvaci2014,Baylis2014}, whose coexistence benefits from common transmit platform, and in turn moving away from independent systems \cite{Huang2015, Griffiths2015, Mealey1963}. In order to enable usage or sharing of spectrum resources and platform hardware, a dual-function radar communications (DFRC) system utilizing waveform diversity in tandem with amplitude/phase control of the radar beam was introduced in \cite{Blunt2010, Ciuonzo2015, Guerci2015, Aubry2014, Aubry2015,Li2016}, where radar is considered as the primary function and presents itself as a system of opportunity to secondary communication functions. In dual-function paradigm, identical signals, same carrier frequency and bandwidth, and common antenna array are deployed to fulfill the objectives of both radar and communication operations. In DFRC system, secondary communications strive to embed a sequence of binary data $b_1, \ldots, b_K$ during each radar pulse which can be achieved through scaling or modulation of either the radar beam or the radar waveform or both. This signal embedding, however, should be accomplished with no, or  minimum, alterations to primary radar function, whether it is detection, tracking, or estimation.

One signaling strategy for embedding information into the radar pulsed emissions uses sidelobe amplitude modulation (AM) and changes the sidelobe level (SLL), according to the information message, towards the intended communication user direction \cite{Hassanien2015}. In lieu of AM, a coherent phase-modulation (PM)-based method was proposed in \cite{Hassanien2015a, Hassanien2016a} to embed one symbol into the radar emission by controlling the phase of complex transmit array pattern. The benefits of decomposing the radar pulse into different waveforms were demonstrated in \cite{Hassanien2016a} where a communication symbol is embedded as a phase rotation between a pair of transmitted orthogonal waveforms. A multi-waveform amplitude shift keying (ASK) strategy was introduced in \cite{Hassanien2016} to embed one binary bit with each orthogonal waveform through bilevel sidelobe control.

Similar to the offering of multi-waveforms, the dual functions of the radar communications system can be improved by properly utilizing the multi-sensor transmit/receive array configurations \cite{McCormick2017, McCormick2017a, Liu2017}. Although the nominal array configuration for existing DFRC systems is uniform and of fixed-structured, it is not necessarily optimum in every sense, and ignores the additional degrees of freedom (DoFs) provided by the flexibility in configuring the antenna array \cite{Pal2010, Nai2009, Joshi2009, Wang2014a, Amin2016}. Non-uniform Sparse arrays have attracted increased attention in multi-sensor transmit/receive systems as an effective solution to reduce the system's complexity and cost, yet retain desired performance \cite{Sanayei2004, Xiao2017, Xiao2017a}. Sparse array design is often cast as optimally placing a given number of antennas on a larger number of possible uniform grid points. In so doing, we are able to span large aperture without introducing unwanted high sidelobes, thus improving spatial resolution. Sparse array design becomes an ``antenna selection'' problem when the number of antennas is equal to the number of grid points, but with fewer RF units. In this case, antenna selection amounts, in essence, to assigning antennas to RF units. In transmit antenna selection, the number of expensive RF chains, which consist of digital-to-analog converter, up-converter, filters and power amplifiers, is smaller than the number of available transmit antenna elements \cite{Molisch2004, Mehta2012}. Thus sparse arrays can, undoubtedly, alleviate pressures on the resource management and efficiency requirements on power amplifiers.

It had been clearly documented in recent papers \cite{Fuchs2012, Nongpiur2014, Wang2014} that the performance of optimum sparse array beamformer is dependent on both array configuration and beamforming weights. In this paper, we add to DFRC system paradigm by introducing a new co-existence approach based on antenna selection. Specifically, we examine the problem of sparse array beampattern synthesis with a fixed number of transmit antennas under the framework of dual functional system design. Three different types of DFRC systems are considered, namely, a system that implements (a) single sparse array with one set of weights, i.e.,  a single beamformer; (b)  single sparse array but with different beamformers; (c) multiple sparse but complementary arrays with different beamformers. The latter case is referred to as ``shared aperture'', where the combined sparse arrays span the given system aperture. The main advantages of utilizing sparse arrays in DFRC systems are manifested by simulation results in the suppression of cross-interference between the two functions and improved hardware efficiency.

The novelty of this paper is summarized as follows:
\begin{itemize}
\item 
We utilize array configurations as additional spatial DoFs to facilitate the co-existence of dual-function radar communications and improve the communication performance without sacrificing radar functions.
\item
We solve the new problem of sparse array beampattern synthesis within the framework of dual function system design and propose a method to enhance the robustness of proposed antenna selection algorithm against initial search point.
\item
We consider dual-functional system platforms equipped with different beamformers associated with shared aperture sparse arrays.

\end{itemize}

The rest of the paper is organized as follows. We provide the system configuration and signal model of the DFRC system with antenna selection network in section \ref{sec:model}. The sparse transmit array design under the framework of common array and single beamformer for two functions is investigated in section \ref{sec:sharearray}. We examine the common sparse array design associated with different beamformers for two functions in section \ref{sec:sharearray1}. The design of sparse arrays for radar and communications under the framework of shared aperture is delineated in section \ref{sec:shareaperture}. Simulation results are provided in section \ref{sec:simulations}. Section \ref{sec:conclusions} summarizes the work of this paper.

\section{System Configuration and Signal Model}
\label{sec:model}

\begin{figure}[!ht]
  \centering
    \includegraphics[trim = {6cm 2cm 6cm 3cm}, scale=0.5]{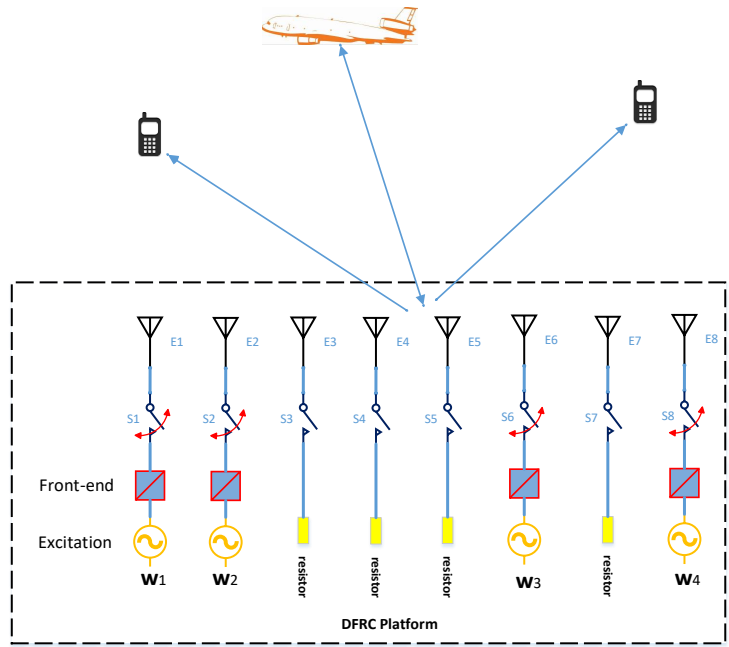}
    \caption{Joint platform of dual function radar communications with antenna selection network.}
\label{fig_1}
\end{figure}

We consider a joint radar communications platform equipped with a reconfigurable transmit antenna array through an antenna selection network, as shown in Fig. \ref{fig_1}. There are $K$ transmit antennas uniformly spaced with an inter-element spacing of $d$ and $M$ front-ends available for waveform transmitting. The antenna selection network comprises $K$ RF switches which connect/disconnect antennas with front-ends. Suppose a transmit array is configured with $M$ selected antennas located at $p_md, m=1,\ldots,M$ with $p_m \in \mathbb{N}$ being non-negative integers. The radar receiver, on the other hand, employs an array of $N$ receive antennas with an arbitrary linear configuration. Without loss of generality, a single-element communication receiver is assumed to be located in the far field at direction $\theta_c$, which is known to the transmitter. Let $\Psi_l(t), l=1, \ldots, L$ be a set of $L$ orthogonal waveforms, each occupying the same bandwidth. In other words, the spectral contents of all waveforms fully overlap in the frequency domain. Each waveform is normalized to have unit power, i.e., $\int_{T} |\Psi_l(t)|^2 dt = 1$, with $T$ and $t$ denoting the radar pulse duration and the fast time index, respectively. It is further assumed that the orthogonality condition $\int_T \Psi_l(t) \Psi_{l'}^*(t) dt = 0$ is satisfied for $l \neq l'$, where $()^*$ stands for the complex conjugate.

Let $\bd s(t;\tau)$ be the $M \times 1$ baseband transmit signal vector during the $\tau$th radar pulse. Assume that $Q$ far-field targets at directions $\theta_q, q = 1, \ldots, Q$, located within the radar main beam, are observed in the background of strong clutter and interference. The $N\times 1$ baseband representation of the signals at the output of the radar receive antenna array is given by,
\begin{equation}
\label{eq:expre_radarrec}
\bd x(t;\tau) = \sum_{q=1}^Q \beta_q(\tau) \bd s^H(t;\tau) \bd a(\theta_q) \bd b(\theta_q) + \bd n(t;\tau),
\end{equation}
where $\beta_q(\tau)$ is the $q$th target reflection coefficient which obeys the Swerling-II target model \cite{Hassanien2015}, i.e., they remain constant during the entire pulse duration, but vary independently from pulse to pulse. The vector $\bd a(\theta)$ is the steering vector of the transmitting array, defined as,
\begin{equation}
\bd a(\theta) = [e^{jk_0 p_1d \sin\theta}, \ldots, e^{jk_0 p_Md \sin\theta}]^T,
\end{equation}
where $k_0=2\pi /\lambda$ is the wavenumber. The steering vector of the receiving array, $\bd b(\theta)$, can be defined in a similar way as $\bd a(\theta)$. The vector $\bd n(t; \tau)$ is of $N \times 1$ dimension, representing the unwanted clutter, interference and white noise in the $\tau$th radar pulse.

The baseband signal received by the communication antenna can be expressed as,
\begin{equation}
\label{eq:expre_communication}
x_c(t;\tau) = \beta_c(\tau) \bd s^H(t;\tau) \bd a(\theta_c)  + n_c(t;\tau),
\end{equation}
where $\beta_c(\tau)$ is the channel coefficient of the received signal that summarizes the propagation environment between the transmit array and the communication receiver during the $\tau$th pulse, and $\bd a(\theta_c)$ is the steering vector of the transmit array toward the communication direction $\theta_c$. In addition, $n_c(t;\tau)$ is the noise signal. We can observe from Eqs. (\ref{eq:expre_radarrec}) and (\ref{eq:expre_communication}) that the array configuration, via the non-uniform inter-element spacing, plays a pivotal role in determining the beamforming performance, which can be considered as an additional DoF designated for performance enhancement. We elaborate on the design of sparse arrays for DFRC systems in the following sections, under different constraints of common and shared system resources.

\section{Design of Common Array with Single Beamformer for DFRC Systems}
\label{sec:sharearray}

This section considers a common transmit array with a single beamformer for both radar and communication functions, and examines the sparse array design for different signaling strategies in phased array radar.

\subsection{Revisit of Different Signaling Schemes}
\label{subsec:revisit}

The general expression of transmit signal for the phased-array radar can be written as,
\begin{equation}
\label{eq:expre_st}
\bd s(t; \tau) = \Psi(t) \bd w(\tau),
\end{equation}
where the $M \times 1$ beamforming weight vector $\bd w(\tau), \|\bd w(\tau)\|_2=1$ is varied in PRIs and required to satisfy a certain desired transmit power radiation pattern. Substituting Eq. (\ref{eq:expre_st}) into Eq. (\ref{eq:expre_communication}), the transmit beamforming complex gain towards the communication receiver becomes $\bd w^H \bd a(\theta_c)  = G_c e^{j \phi_c}$ with $G_c$ and $\phi_c$ denoting the magnitude and phase, respectively. During each radar pulse, communications can be simultaneously achieved with radar functions by embedding information into either the amplitude $G_c$ or the phase $\phi_c$. The associated methods amount to radar beam modulations, and are termed amplitude modulation (AM)- and phase modulation (PM)-based methods, respectively.

The AM-based method is to embed information into the radar emission via modulating the complex gain amplitude $G_c$ towards the intended communication direction. To satisfy the primary radar operation requirements, the radar mainlobe is kept unchanged during the entire CPI. Thus, the AM-based method can only enable information delivery to a communication receiver located within the sidelobe region. Accordingly, a narrower radar mainlobe implies a wider angular sector where the communication function can take place. The $N_b$ information bits are mapped into a dictionary of $K=2^{N_b}$ sidelobe levels (SLLs) denoted as $G_c \in \{\Delta_1, \ldots, \Delta_K\}$. Therefore, the implementation of this method requires a single radar waveform $\Psi(t)$ and $K$ beamforming weight vectors associated with distinct SLLs. The PM-based method, on the other hand, embeds information by controlling the phase $\phi_c$ of the beam radiated towards the communication receiver. The $N_b$ sequence of binary bits are mapped into a dictionary of $K$ phase symbols denoted as $\phi_c \in \{\phi_1, \ldots, \phi_K\}$. For coherent communications, this method amounts to using one waveform and $K$ beamforming weights associated to deliver $K$ distinct phases. 

Matched filtering the received communication signal in Eq. (\ref{eq:expre_communication}) with the waveform $\Psi(t)$ yields,
\begin{equation}
\label{eq:expre_yc}
y_c(\tau) = \int_T x_c(t,\tau) \Psi(t) dt = \beta_c(\tau) G_c e^{j\phi_c} + n_c(\tau),
\end{equation}
where $n_c(\tau)=\int_T n_c(t;\tau) \Psi(t) dt$ is the additive noise term after integration. The embedded communication symbol can be estimated as,
\begin{equation}
\label{eq:symbol_estimate}
\hat{G}_c = \left|\frac{y_c(\tau)}{\beta_c(\tau)}\right| \quad \t{and} \quad \hat{\phi}_c = \t{angle}(y_c(\tau))-\t{angle}(\beta_c(\tau)),
\end{equation}
where $|\cdot|$ stands for absolute value and $\t{angle}(\cdot)$ is the angle of the argument. The actual embedded binary message can be decoded by comparing the complex gain estimate with the pre-defined $K$-dimensional dictionary. Both the AM and PM methods can achieve a data rate of $R_b = N_b \cdot \t{PRF}$ in bits per second, with PRF denoting the pulse repetition frequency. The hybrid amplitude phase modulation method, such as quadrature amplitude modulation (QAM), can certainly achieve a doubled data rate by choosing the amplitude $\Delta_k$ and the phase $\phi_k$ from the respective pre-defined dictionary $G_c$ and $\phi_c$, compared with single AM or PM beam modulation.

In addition the above methods, two other signaling strategies for information embedding using waveform diversity in tandem with beamforming gain control were proposed in \cite{Hassanien2016}. The multi-waveform two-level ASK (2ASK) and binary phase shift keying (BPSK) methods employ multiple waveforms in conjunction with two transmit beamforming weight vectors, denoted as $\bd w_H$ and $\bd w_L$ ($\bd w_P$ and $\bd w_Q$ for BPSK). In order to embed $N_b$ bits per radar pulse, $N_b$ orthogonal waveforms are simultaneously transmitted with each waveform delivering one information bit to the communication receiver. During each radar pulse, the waveform $\Psi_n(t), n=1, \ldots, N_b$ is radiated either via $\bd w_H$ ($\bd w_P$) for $b_n = 1$ or $\bd w_L$($\bd w_Q$) when $b_n=0$. Accordingly, the transmit signals in Eq. (\ref{eq:expre_st}) can be rewritten as,
\begin{equation}
\bd s(t) = \sum_{n=1}^{N_b} \sqrt{\frac{P_t}{N_b}} \left( b_n \bd w_i + (1-b_n) \bd w_j \right) \Psi_n(t), i=H,P; j=L,Q.
\end{equation}
Here, the power budget $P_t$ is assumed to be equally distributed among the $N_b$ orthogonal waveforms. Matched filtering the received signal with each orthogonal waveform yields the $N_b \times 1$ data vector $y_{c,n}, n=1, \ldots, N_b$, defined as
\begin{equation}
y_{c,n}(\tau) = \sqrt{\frac{P_t}{N_b}} \beta_c(\tau) [(1-b_n)\bd w_i + b_n \bd w_j]^H \bd a(\theta_c) + n_{c,n}(\tau), n=1, \ldots, N_b, i=H,P; j=L,Q.
\end{equation}
where $n_{c,n}(\tau) = \int_T n_c(t;\tau) \Psi_n(t) dt, n=1, \ldots, N_b$. The embedded communication symbol can then be estimated according to Eq. (\ref{eq:symbol_estimate}). Clearly, in this case, the data rate is $N_b \times \t{PRF}$ bits per second.

\subsection{Design of Common Array with Single Beamformer for DFRC Systems}

For a given transmit array configuration, the design of the beamforming weight vector $\bd w$ embedding the $k$th communication symbol $\Delta_k e^{j \phi_k}$ can be formulated in two different cases. In the case of desired focused beampattern for the radar function, the weight vector $\bd w$ is designed to maintain the power radiation in the sidelobe region $\bar{\Theta}$ under the specified level $\rho$, while maintaining a unit gain towards the radar target direction $\theta_t$. The problem formulation can be written as,
\begin{eqnarray}
\label{eq:expre_weightradar1}
\textbf{Focused Pattern:}\quad \min_{\bd w, \alpha} && \alpha, \\
\t{subject \; to} && \bd w^H \bd a(\theta_t) = e^{j \mu(\theta_t)}, \nn\\
                  && \left| \bd w^H \bd a(\theta_k) \right| \leq \rho + \alpha, \; \theta_k \in \bar{\Theta}, k = 1, \ldots, L_s \nn\\
                  && \bd w^H \bd a(\theta_c) = \Delta_k e^{j\phi_k}, \; k=1, \ldots, K,  \nn
\end{eqnarray}
where $\theta_k, k=1,\ldots,L_s$ are $L_s$ samples of the sidelobe region $\bar{\Theta}$, and $\mu(\theta_t)$ denotes the phase of unit complex gain. Here, $\alpha$ is an auxiliary variable for controlling sidelobe levels and its maximum value is 0. In the case of desired flat-top beampattern for the radar function, the main function of $\bd w$ is to concentrate the transmit power within a certain angular sector $\Theta=[\theta_{\t{min}}, \theta_{\t{max}}]$. The mainlobe ripples are required to be less than a specified level $\epsilon$ and the power radiation level corresponding to the sidelobe region $\bar{\Theta}$ is well-controlled under the specified value $\rho$. This problem can be formulated as,
\begin{eqnarray}
\label{eq:expre_weightradar}
\textbf{Flat-top Pattern:} \quad \min_{\bd w, \alpha} && \alpha, \\
\t{subject \; to} && \left| \bd w^H \bd a(\theta_i) - e^{j \mu(\theta_i)}\right| \leq \epsilon, \; \theta_i \in \Theta, i = 1, \ldots, L_m, \nn\\
                  && \left| \bd w^H \bd a(\theta_k) \right| \leq \rho + \alpha, \; \theta_k \in \bar{\Theta}, k = 1, \ldots, L_s \nn\\
                  && \bd w^H \bd a(\theta_c) = \Delta_k e^{j\phi_k}, \; k=1, \ldots, K,  \nn
\end{eqnarray}
where $\theta_i, i=1,\ldots,L_m$ are samples of the mainlobe region $\Theta$, and $\mu(\theta_i), i=1, \ldots, L_m$ stands for the mainlobe phase profile. Different from the formulation in \cite{Hassanien2016a}, the mainlobe phase profile $\mu(\theta_i), i=1, \ldots, L_m$ is adjustable in lie of assuming fixed values. We use the phase profile of the desired beampattern, $e^{j\mu(\theta)}$, as a free parameter in the optimization in order to achieve better approximation to the desired beampattern. In this respect, we utilize an alternating descent method that iteratively shifts between the weight vector $\bd w$ and the phase profile $e^{j\mu(\theta)}$ \cite{Wang2014}. In the $l+1$th iteration, we compute the weight vector $\bd w^{(l+1)}$ by solving Eq. (\ref{eq:expre_weightradar}) based on the phase profile $e^{j \mu^{(l)}(\theta)}$ from the previous $l$th iteration. Then, we update the phase profile as follows,
\begin{equation}
\label{eq:expre_phasemain}
e^{j \mu^{(l+1)}(\theta_i)} = \frac{\bd w^{(l+1)H}a(\theta_i)}{|\bd w^{(l+1)H}a(\theta_i)|}, i=1, \ldots, L_m, \theta_i \in \Theta.
\end{equation}
It was proved in \cite{Wang2014} that the combination of phase profile updating formula in Eq. (\ref{eq:expre_phasemain}) and alternating descent method is capable of converging to the ideal power pattern synthesis within the mainlobe angular region, that is $-\epsilon \leq |\bd w^H \bd a(\theta_i)| - 1 \leq \epsilon, i=1, \ldots, L_m$ with $\epsilon$ denoting the maximum allowable ripple.

The beamforming weight design in Eqs. (\ref{eq:expre_weightradar1}) and (\ref{eq:expre_weightradar}) often assumes a uniform transmit array configuration with $M$ antennas. It is difficult to design a large-aperture uniform linear array (ULA) due to increased number of antennas and associated hardware cost, which may, in turn, limit the angular region where the communication function takes place. The optimally designed sparse array is capable of achieving high angular resolution without introducing unwanted high sidelobes. 

The sparse array transmit pattern synthesis considered in this work involves the integrated design of sparse array configuration and the corresponding beamforming weights. Define an antenna selection vector $\bd r \in \{0,1\}^M$ with entry ``1'' denoting the respective antenna selected and entry ``0'' discarded. The positions of ``1'' entries in the vector $\bd r$ determine the transmit array configuration. The beamforming weight vector $\bd w$ is mandated to exhibit the same sparse structure as the selection vector $\bd r$. The design of a common sparse array with a single beamformer should satisfy both radar and communication functions. Taking the focused beampattern synthesis in Eq. (\ref{eq:expre_weightradar1}) as an example, the sparse array design for dual-functional systems can be formulated as follows,
\begin{subequations}
\begin{eqnarray}
\label{eq:expre_designwo_2}
\min_{\bd w, \bd r, \alpha} && \alpha, \\
\t{subject \; to} && \bd w^H \bd a(\theta_t) = e^{j \mu(\theta_t)}, \nn\\
                  && \left| \bd w^H \bd a(\theta_l) \right| \leq \rho + \alpha, \; \theta_l \in \bar{\Theta}, l = 1, \ldots, L_s \nn \\
                  \label{eq:expre_designwo_2_con1}
                  && \bd w^H \bd a(\theta_c) = \Delta_k e^{j\phi_k}, \\
                  \label{eq:expre_designwo_2_con2}
                  && |w_m|^2 \leq r_m, m = 1, \ldots, M \\
                  \label{eq:expre_designwo_2_con3}
                  && \bd r \in \{0, 1\}^M, \; \bd 1^T \bd r = M,
\end{eqnarray}
\end{subequations}
where the constraint in Eq. (\ref{eq:expre_designwo_2_con1}) is used to embed the communication symbol $\Delta_k e^{j\phi_k}$. The constraint in Eq. (\ref{eq:expre_designwo_2_con2}) is imposed to couple the two variables $\bd w$ and $\bd r$, and promotes the same sparsity of the beamforming weight vector $\bd w$ as the selection vector $\bd r$. The constraint in Eq. (\ref{eq:expre_designwo_2_con3}) restrains the number of selected antennas to be exactly $M$. 

The problem in Eq. (\ref{eq:expre_designwo_2}) belongs to notorious combinatorial optimization. In order to eliminate the boolean constraint in Eq. (\ref{eq:expre_designwo_2_con3}), the formulation in Eq. (\ref{eq:expre_designwo_2}) can be rewritten as,
\begin{eqnarray}
\label{eq:expre_designwo_1}
\min_{\bd w, \bd r, \alpha} && \alpha + \gamma \bd r^T (\bd 1- \bd r), \\
\t{subject \; to} && \bd w^H \bd a(\theta_t) = e^{j \mu(\theta_t)}, \nn\\
                  && \left| \bd w^H \bd a(\theta_l) \right| \leq \rho + \alpha, \; \theta_l \in \bar{\Theta}, l = 1, \ldots, L_s \nn \\
                  && \bd w^H \bd a(\theta_c) = \Delta_k e^{j\phi_k}, \nn\\
                  && |w_m|^2 \leq r_m, m = 1, \ldots, M \nn\\
                  && 0 \leq \bd r \leq 1, \; \bd 1^T \bd r = M. \nn
\end{eqnarray}
Here, the boolean constraint on $\bd r$ is tantamount to the combination of the box constraint $0 \leq \bd r \leq 1$ and the second part of the objective function $\min_{\bd r} \; \bd r^T (\bd 1- \bd r)$. Additionally, $\gamma$ represents a trade-off parameter which compromises between the peak sidelobe level (PSL) and the sparseness of the selection vector $\bd r$. The two formulations in Eqs. (\ref{eq:expre_designwo_2}) and (\ref{eq:expre_designwo_1}) become equivalent when $\gamma$ tends to infinity \cite{Tuy1998, Horst2000}.

The second part of the objective function $\bd r^T(\bd 1 - \bd r)$ is concave, and thus it is difficult to minimize Eq. (\ref{eq:expre_designwo_1}) directly. A sequential convex programming (SCP) based on iteratively linearizing the second concave function is utilized to reformulate the non-convex problem to a series of convex subproblems, each of which can be optimally solved using convex programming \cite{Boyd2004,Fazel2003}. The integrated sparse array design and beampattern synthesis in the (k+1)th iteration can be formulated based on the solution from the kth iteration $\bd r^{(k)}$ as,
\begin{eqnarray}
\label{eq:expre_designwo_11}
\min_{\bd w, \bd r, \alpha} && \alpha + \gamma [\bd r^T (\bd 1- 2\bd r^{(k)}) + \bd r^{(k)T} \bd r^{(k)}], \\
\t{subject \; to} && \bd w^H \bd a(\theta_t) = e^{j \mu(\theta_t)}, \nn\\
                  && \left| \bd w^H \bd a(\theta_l) \right| \leq \rho + \alpha, \; \theta_l \in \bar{\Theta}, l = 1, \ldots, L_s \nn \\
                  && \bd w^H \bd a(\theta_c) = \Delta_k e^{j\phi_k}, \nn\\
                  && |w_m|^2 \leq r_m, m = 1, \ldots, M \nn\\
                  && 0 \leq \bd r \leq 1, \; \bd 1^T \bd r = M. \nn
\end{eqnarray}
Eq. (\ref{eq:expre_designwo_11}) can be formulated into a standard second order cone programming (SOCP) and effectively solved by various types of software packages, such as CVX \cite{Grant2008}. Note that Eq. (\ref{eq:expre_designwo_11}) is a convex relaxation of the original combinatorial optimization problem in Eq. (\ref{eq:expre_designwo_2}), thus the SCP is a local heuristic and the obtained solution is a sub-optimum local minimizer. In order to find a sufficiently good sub-optimum sparse array, the typical remedy is to initialize the algorithm with several feasible points $\bd r^{(0)}$ and find the one with the minimum objective value over the different runs. To accelerate the convergence rate of the algorithm and increase its robustness against initial search points, we propose a new method to update the selection vector $\bd r$ iteratively. The updating rule of the selection vector in the $k+1$th iteration is expressed as,
\begin{equation}
\label{eq:expre_rupdate}
\bd r^{(k+1)} = \bd r^{(k)} \oplus \Delta \bd r,
\end{equation}
where $\oplus$ denotes modulus-two addition and $\Delta \bd r \in \{0,1\}^K$ is a $K$-dimensional vector with all zeros except for two entries with value one. The positions $p_i, i=1,2$ of the two ``one'' entries are determined by $\bd w^{(k+1)}$ and $\bd r^{(k)}$ together as follows,
\begin{eqnarray}
\label{eq:expre_rupdate1}
p_1 &=& \rm argmax_n \{ | w^{(k+1)}_n | : r^{(k)}_n=0 \}, \\
p_2 &=& \rm argmin_n \{ | w^{(k+1)}_n | : r^{(k)}_n=1 \}. \nn 
\end{eqnarray}
The integrated design of sparse array configuration $\bd r$ and the associated beamforming weight $\bd w$ comprises two main stages. In the first stage, a sparse array $\bd r$ and the associated beamformer $\bd w_1$ are obtained for embedding the first communication symbol from Eq. (\ref{eq:expre_designwo_11}). In the second stage, all other beamformers $\bd w_k, k=2,\ldots K$ for embedding the remaining symbols based on the obtained sparse array are calculated from Eq. (\ref{eq:expre_weightradar1}). Thus, it is not necessary to perform antenna selection and reconfigure the array at each PRI. Note that an alternating descent algorithm is deployed in both stages to synthesize the desired beampattern by iteratively updating the mainlobe phase profile. The phase profile $\mu(\theta_i), \theta_i \in \Theta$  is initialized according to the following formula,
\begin{equation}
\mu(\theta_t)=1, \t{for focused beam}; \quad \mu(\theta_i) = -2\pi \sin(\theta_i), \theta_i \in \Theta, \t{for flat-top beam}.
\end{equation}
The detailed description of the sparse array design for DFRC systems is summarized in Table \ref{table_1}. To reduce the effect of initial search points, in addition to the proposed updating rule, a new starting point is selected whenever the iteration number exceeds the maximum number $Q$.
\begin{table}[!h]
\caption{The detailed description of sparse array design for DFRC systems}
\label{table_1}
\begin{center}
\begin{tabular}{c|c}
\hline
\hline
Step 0 & Initialize the threshold value $\delta$ and phase profile $\mu(\theta)$; Set the maximum iteration number $Q$;\\
Step 1 & generate a random feasible starting point $\bd r^{(0)}$, \\
Step 2 & \textbf{WHILE:} Run the optimization in Eq. (\ref{eq:expre_designwo_11}) based on $\bd r^{(k)}$ to obtain $\bd w^{(k+1)}$;\\
Step 3 & Update $\bd r^{(k+1)}$ according to Eqs. (\ref{eq:expre_rupdate}) and (\ref{eq:expre_rupdate1}) based on $\bd w^{(k+1)}$ ans $\bd r^{(k)}$;\\ 
Step 3 & Update the phase profile $\mu^{(k+1)}(\theta)$ according to Eq. (\ref{eq:expre_phasemain}).\\
Step 4 & If $\|\bd w^{(k+1)}-\bd w^{(k)}\|_2 \geq \delta$ and $k \leq Q$, set $k=k+1$ and go to Step 2; \\
       & If $\|\bd w^{(k+1)}-\bd w^{(k)}\|_2 \leq \delta$, store the obtained sparse beamformer and terminate. \\
Step 5 & If $\|\bd w^{(k+1)}-\bd w^{(k)}\|_2 > \delta$ and $k > Q$, go to Step 1. \textbf{END}\\
\hline
\hline
\end{tabular}
\end{center}
\end{table}

\section{Design of Common Array with Multiple Beamformers for DFRC Systems}
\label{sec:sharearray1}

\begin{figure}[!ht]
  \centering
    \includegraphics[trim = {6cm 2cm 6cm 3cm}, scale=0.5]{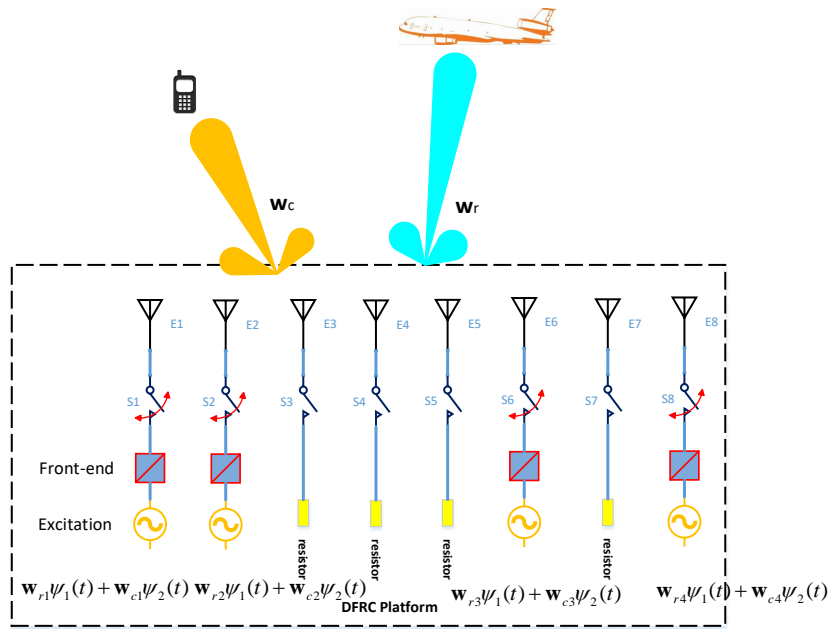}
    \caption{Joint platform of DFRC systems with a common sparse array and multiple beamformers.}
\label{fig_multi}
\end{figure}

Now, we assume that both radar and communication functions deploy the same sparse transmit array while being associated with different beamformers. This strategy is illustrated in Fig. \ref{fig_multi}, where $\bd w_r$ and $\bd w_c$ are weight vectors corresponding to two beamformers for radar and communications, respectively. The cross-interference between different beamformers can be mitigated by utilizing orthogonal waveforms and spatial filtering. The composite transmit signal is written as,
\begin{equation}
\bd s(t;\tau) = \bd w_r \Psi_1(t) + \bd w_c \Psi_2(t),
\end{equation}
Matched filtering the radar received signal in Eq. (\ref{eq:expre_radarrec}) with the waveform $\Psi_1(t)$ yields,
\begin{equation}
\bd x(\tau) = \sum_{q=1}^Q \beta_q(\tau) \bd w_r^H \bd a(\theta_q) \bd b(\theta_q) + \bd n(\tau).
\end{equation}
Similarly, matched filtering the communication received signal in Eq. (\ref{eq:expre_communication}) with the waveform $\Psi_2(t)$ yields,
\begin{equation}
x_c(\tau) = \beta_c(\tau) \bd w_c^H \bd a(\theta_c)  + n_c(\tau).
\end{equation}
Thus, the radar and communication functions do not interfere with each other by utilizing the waveform diversity. As such, the beamformer $\bd w_r$ is responsible to both synthesize the desired beampattern for radar applications and mitigate the interference to communications caused by the radar function. On the other hand, the beamformer $\bd w_c$ is utilized to provide directive gain towards the communication receiver. The common sparse array with two respective beamformers for radar and communications can be designed as follows,
\begin{subequations}
\begin{eqnarray}
\label{eq:expre_designwo11}
\min_{\bd w_r, \bd w_c, \alpha} && \alpha + \gamma \bd r^T (\bd 1- \bd r), \\
\t{subject \; to} && \bd w_r^H \bd a(\theta_t) = e^{j \mu(\theta_t)}, \nn\\
                  && \left| \bd w_r^H \bd a(\theta_l) \right| \leq \rho + \alpha, \; \theta_l \in \bar{\Theta}, l = 1, \ldots, L_s \nn \\
                  \label{eq:expre_designwo11_con1}
                  && \left| \bd w_c^H \bd a(\theta_l) \right| \leq \rho + \alpha, \; \theta_l \in \bar{\Theta}_c, l = 1, \ldots, L_c \\
                  \label{eq:expre_designwo11_con2}
                  && \bd w_r^H \bd a(\theta_c) = 0, \\
                  \label{eq:expre_designwo11_con5}
                  && \bd w_c^H \bd a(\theta_c) = 1; \\
                  \label{eq:expre_designwo11_con3}
                  && |w_{i,m}|^2 \leq r_m, m = 1, \ldots, M, i=\{r,c\} \\
                  \label{eq:expre_designwo11_con4}
                  && 0 \leq \bd r \leq 1, \; \bd 1^T \bd r = M,
\end{eqnarray}
\end{subequations}
where $\bar{\Theta}_c$ is the sidelobe angular region pre-defined for communications, which contains the radar mainlobe sector $\Theta$. The constraint in Eq. (\ref{eq:expre_designwo11_con1}) restrains the impact of communications on the radar function to be less than $\rho$. The constraint in Eq. (\ref{eq:expre_designwo11_con2}) imposes orthogonality between the radar beamforming weights and the communication steering vector, such that the radar function does not interfere the communication function. The constraint $\bd w_c^H \bd a(\theta_c) = 1$ provides unit gain towards the communication receiver. The two constraints in Eqs. (\ref{eq:expre_designwo11_con3}) guarantee the common sparse array for two respective beamformers associated with radar and communication functions. We do not restrain the communication beamformer to put a null against the radar target, as the reflected communication signal may be utilized by the radar for detection performance improvement. It is worth noting that communications and radar functions can be implemented independently and concurrently due to different beamformers and orthogonal waveforms. In this respect, it is unnecessary to embed communication symbols into the radar pulses by utilizing the signaling strategies described in section \ref{subsec:revisit}.

\section{Design of Intertwined Subarrays with Shared Aperture For DFRC Systems}

\label{sec:shareaperture}

\begin{figure}[!ht]
  \centering
    \includegraphics[trim = {6cm 2cm 6cm 3cm}, scale=0.5]{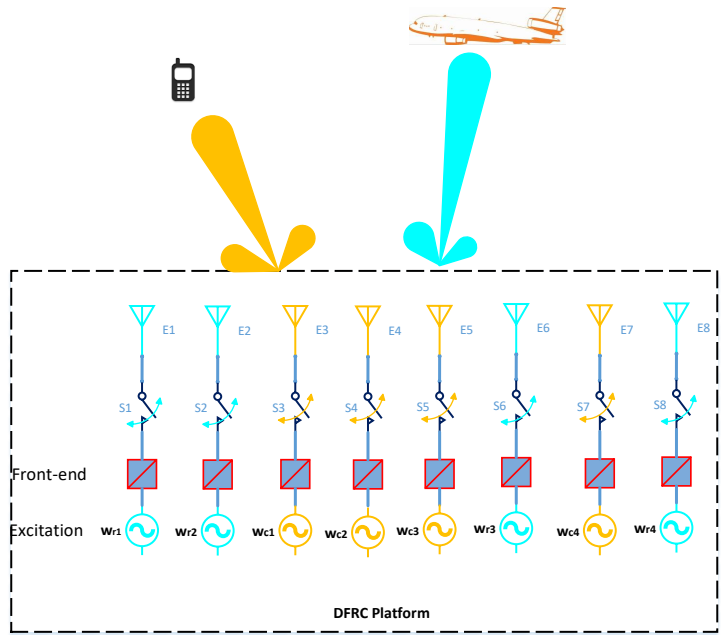}
    \caption{Joint platform of DFRC systems with intertwined subarrays for radar and communications.}
\label{fig_2}
\end{figure}

In this section, we consider a separated antenna deployment strategy which partitions the $K$ available antennas into two sparse subarrays, one for radar and the other for communications. Assume that $M_r$ out of $K$ antennas are allocated for radar and the remaining $M_c=K-M_r$ antennas for communications. The two corresponding subarrays are entwined and together form a contiguous filled aperture. The $M_r$-antenna radar transmit subarray is capable of concentrating the transmit power towards the target and a deep null against the communication receiver to mitigate the cross-interference. On the contrary, the remaining $M_c$-antenna communication subarray is required to point the beam towards the communication receiver and maintains well-controlled sidelobes in other angular regions. The formulation of this problem is similar to Eq. (\ref{eq:expre_designwo11}), with the only difference that the constraints in Eqs. (\ref{eq:expre_designwo11_con3}) and (\ref{eq:expre_designwo11_con4}), respectively, change to
\begin{equation}
\label{eq:expre_designwo1}
|w_{r,m}|^2 \leq r_m, \; |w_{c,m}|^2 \leq 1-r_m, m = 1, \ldots, K
\end{equation}
and
\begin{equation}
0 \leq \bd r \leq 1, \; \bd 1^T \bd r = M_r,
\end{equation}
which implies that the two beamformers exhibit complementary sparse structures, instead of a common structure as per Eq. (\ref{eq:expre_designwo11}). Note that the sparse supports of two beamformers $\bd w_r$ and $\bd w_c$ are interleaved as restricted in Eq. (\ref{eq:expre_designwo1}), and combined beamformer sparse arrays span the entire array aperture. The sparse array configuration remains unchanged with respect to PRI for different communication symbols. Reconfiguration is required when the electromagnetic environment changes, such as the direction of the communication receiver. Various off-the-shelf software packages have been developed to effectively and efficiently solve the above SOCP problem \cite{Domahidi2013, Mattingley2010, Boyd2011}, which facilitates the real-time sparse array reconfiguration in time-varying environment.

\section{Simulations}
\label{sec:simulations}

In the simulations, we consider a radar platform with $K=40$ antennas arranged in a ULA with an inter-element spacing of $0.25 \lambda$. The antenna selection network is deployed to select a subset of $M$ antennas in Examples 1 and 2, and select two entwined subarrays of $20$ antennas in Example 3. We evaluate the performance of sparse arrays in DFRC systems by plotting the power pattern and the bit error rate (BER) curves in different scenarios. 

\subsection{Example 1: Common Array Design with Single Beamformer}

\begin{figure}[!ht]
  \centering
    \includegraphics[trim = {8cm 12cm 6cm 10cm}, scale=0.8]{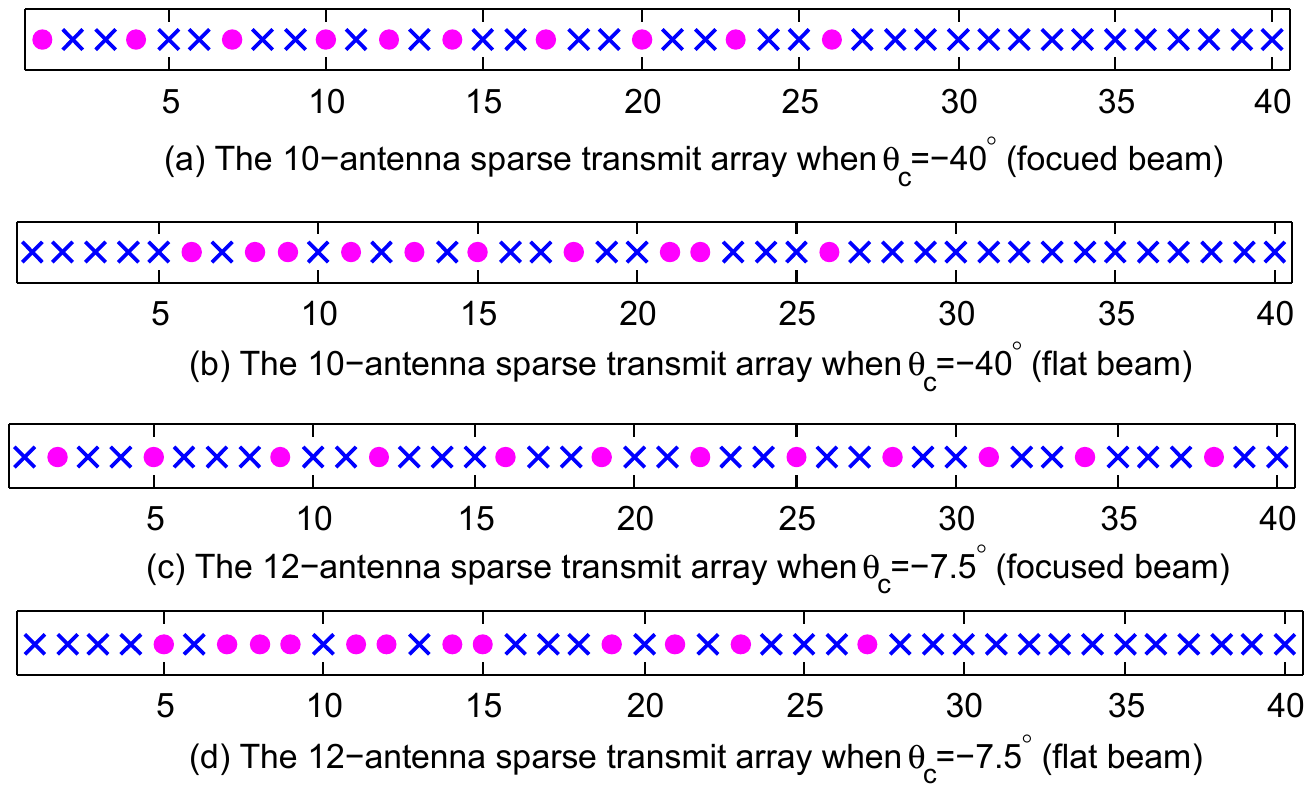}
    \caption{Configurations of proposed sparse arrays: (a) 10-antenna sparse array when $\theta_c=-40^{\circ}$ for focused beam; (b) 10-antenna sparse array when $\theta_c=-40^{\circ}$ for flat-top beam; (c) 12-antenna sparse array when $\theta_c=-7.5^{\circ}$ for focused beam; (d) 12-antenna sparse array when $\theta_c=-7.5^{\circ}$ for flat-top beam. Filled circles represents selected and cross for discarded.}
\label{fig_3}
\end{figure}

\begin{figure}[!ht]
  \centering
    \includegraphics[trim = {6cm 10cm 6cm 9cm}, scale=0.8]{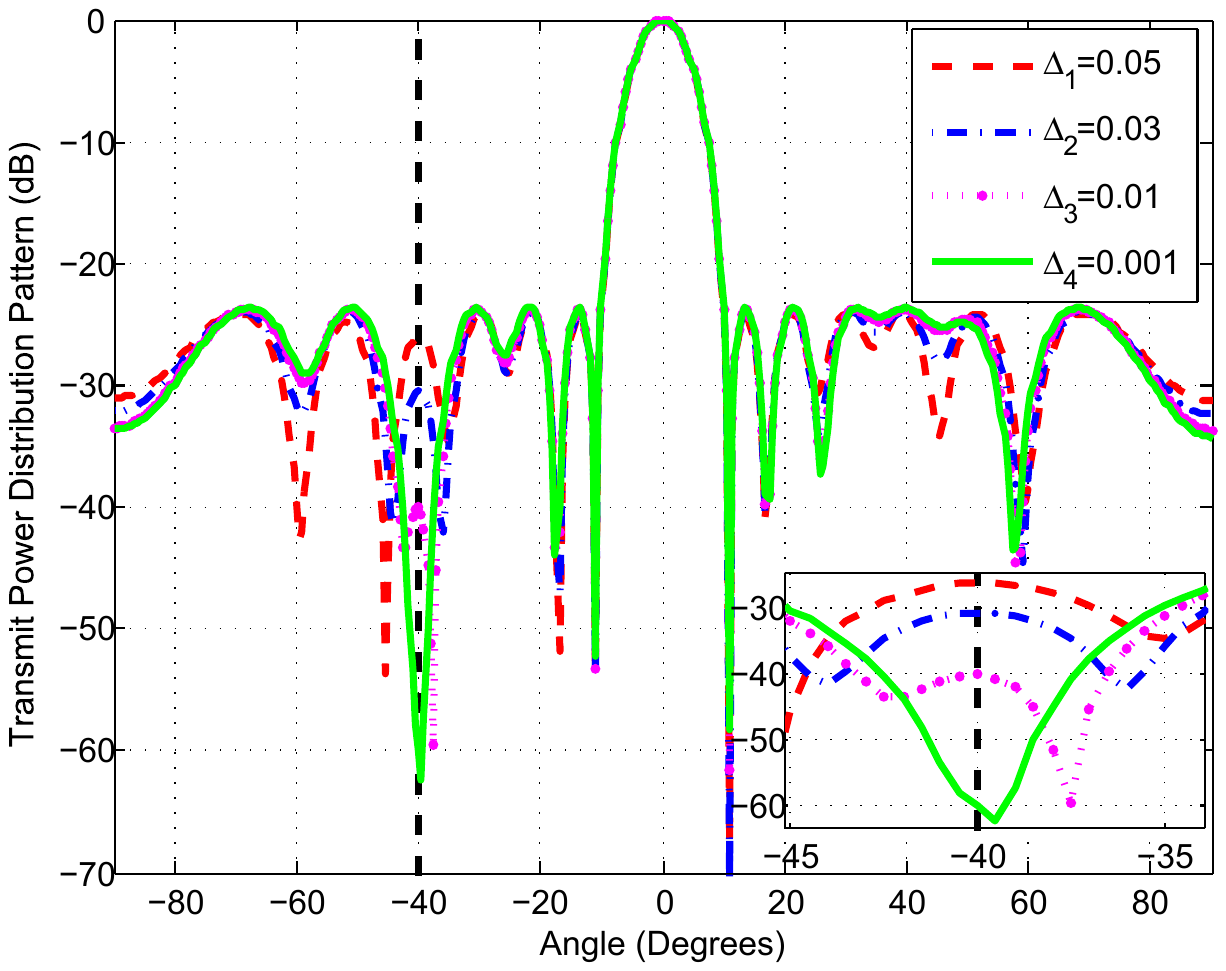}
    \caption{Synthesized focused beampatterns of 10-antenna sparse array (a) for communication symbols $\Delta=\{0.05, 0.03, 0.01, 0.001\}$.}
\label{fig_4}
\end{figure}

\begin{figure}[!ht]
  \centering
    \includegraphics[trim = {6cm 10cm 6cm 9cm}, scale=0.8]{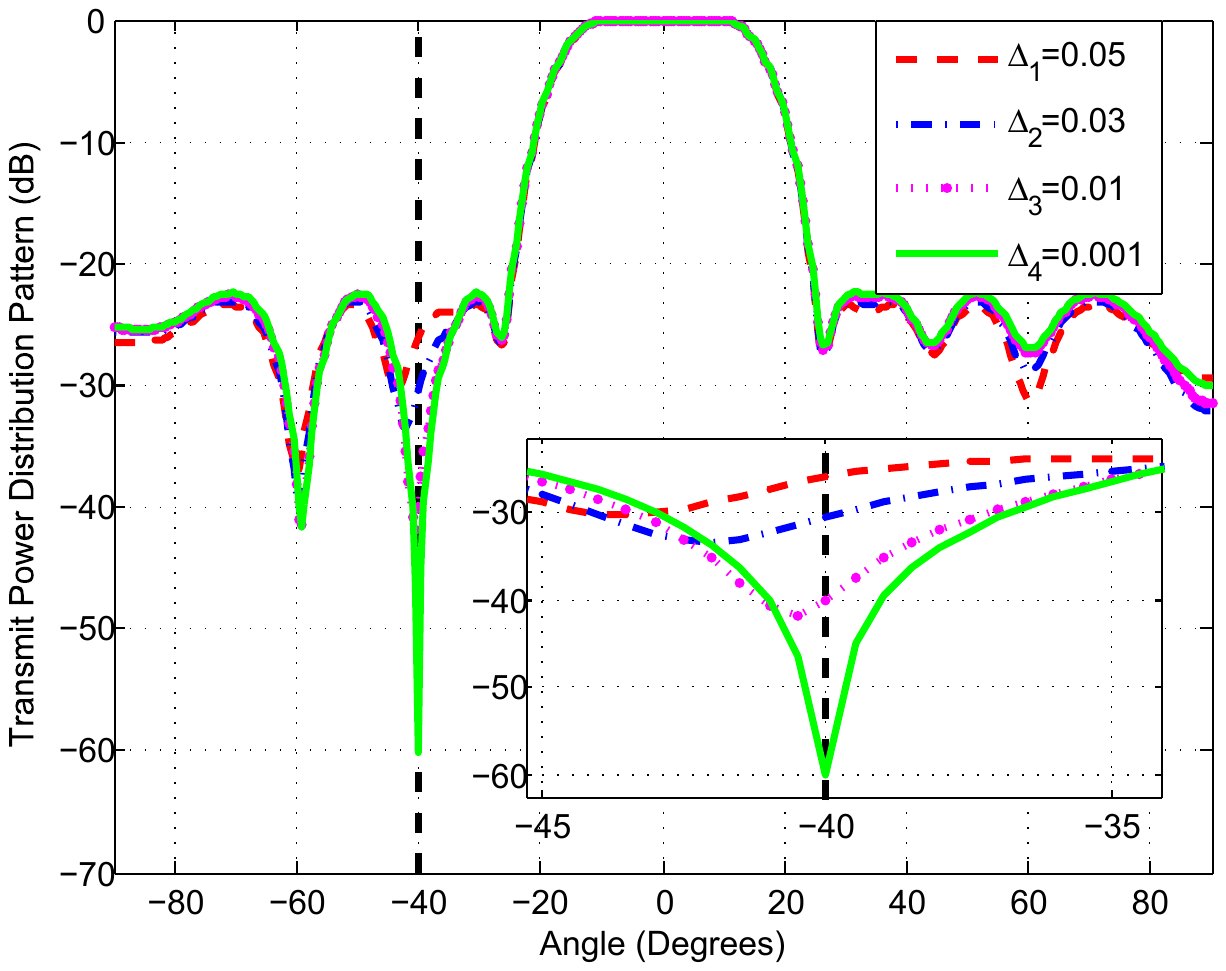}
    \caption{Synthesized flat-top beampatterns of 10-antenna sparse array (b) for communication symbols $\Delta=\{0.05, 0.03, 0.01, 0.001\}$.}
\label{fig_44}
\end{figure}

We first investigate the common sparse array design with a single beamformer for DFRC systems. We assume that the radar target is arriving from the direction $\theta_t=0^{\circ}$ in the focused beam case and within the angular sector $\Theta=[-10^{\circ}, 10^{\circ}]$ in the flat-top beam case. A single communication receiver is assumed at direction $\theta_c=-40^{\circ}$. Two sparse arrays of $10$ antennas are selected for two cases of focused and flat-top beams, respectively. The communication information is embedded during each radar pulse via both the AM and PM signaling schemes. The four communication symbols for the AM and PM modulations are pre-defined as $\Delta = \{0.05, 0.03, 0.01, 0.001\}$ and $\phi_c = \{-\pi/2, 0, \pi/2, \pi\}$, respectively. The two selected $10$-antenna sparse arrays are denoted as array (a) and array (b), and plotted in Fig. \ref{fig_3}. The beampatterns of the sparse arrays (a) and (b) with four modulated SLLs towards the communication receiver are depicted in Figs. \ref{fig_4} and \ref{fig_44}. Clearly, the selected sparse arrays do not utilize the available full aperture, as the communication receiver is located far from the radar target and the beampattern mainlobe is relatively wide. The PSL is set as $-20$dB for both cases and the maximum peak-to-peak mainlobe ripples are set as $0.8$dB. Both the focused and flat-top beampatterns exhibit almost the same shape in the entire angular region excluding the communication receiver direction. The radiation power level of both sparse arrays in the radar sidelobe areas is at least $23$dB lower than the mainlobe.

\begin{figure}[!h]
  \centering
    \includegraphics[trim = {6cm 1cm 6cm 1cm}, scale=0.3]{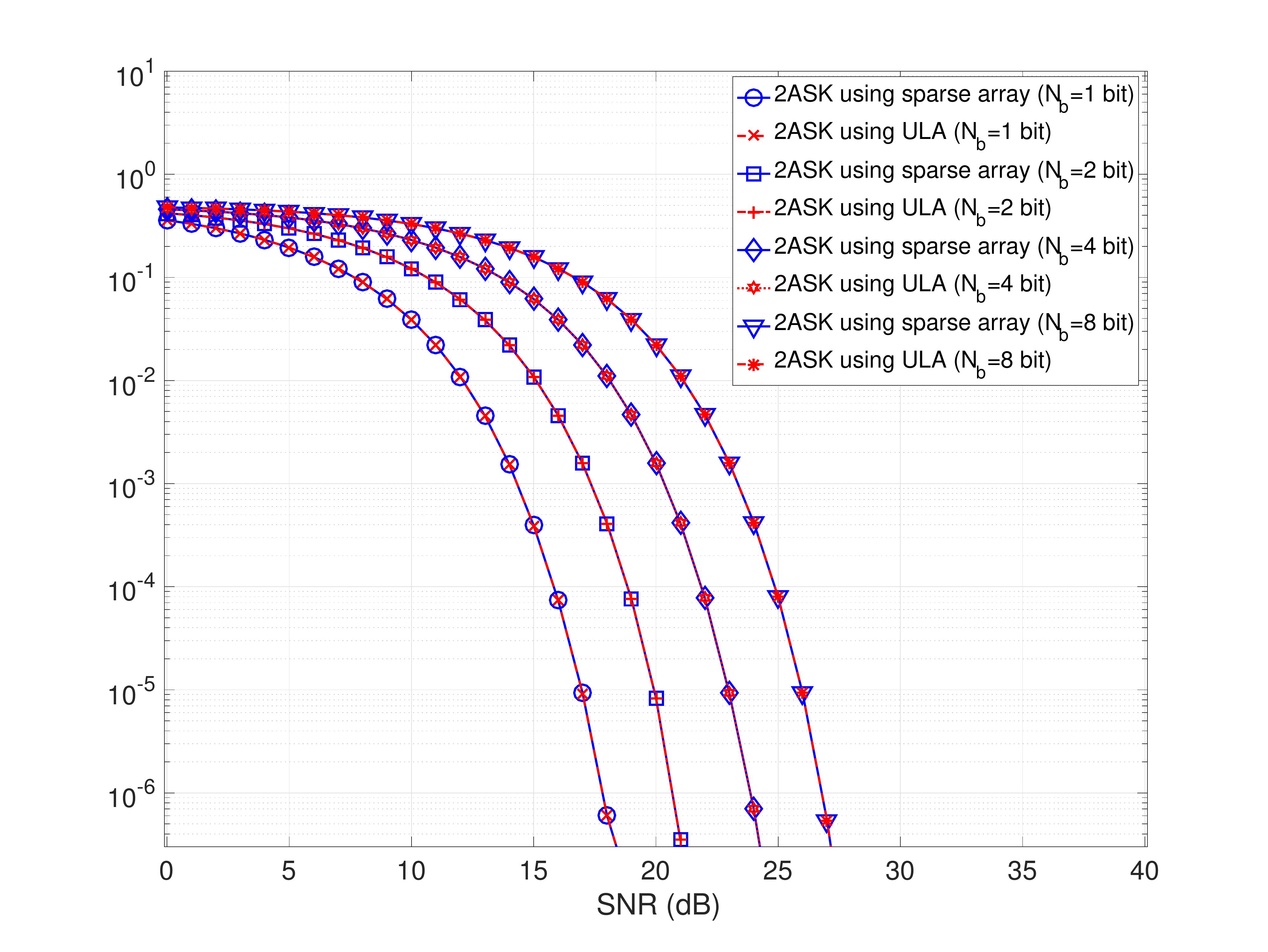}
    \caption{BER versus SNR using 10-antenna sparse array (a) and 10-antenna ULA. Communication receiver locates at direction $\theta_c=-40^\circ$.}
\label{BERfig1}
\end{figure}

Figure~\ref{BERfig1} shows the BER curves versus signal-to-noise ratio (SNR) for variable data rates using the sparse array shown in Fig.~\ref{fig_3} (a) as well as a 10-antenna ULA with a half-wavelength inter-element spacing. The data rates used are 1, 2, 4, and 8 bits per pulse. The SNR is defined as $10\log_{10}(P_t/\sigma_n^2)$ with $P_t$ and $\sigma_n^2$ denoting the transmit power and noise power, respectively. Information embedding is performed using 2ASK and orthogonal waveforms, i.e., eight orthogonal waveforms are used for the case of 8 bits per pulse. The figure shows that all BER curves decrease as the SNR increases. However, since the total power budget $P_t$ is fixed, the BER curve associated with the 2 bits per pulse is shifted by 3dB to the right on the SNR axis as compared to the BER curve associated with the 1 bit per pulse case. This is attributed to the fact that the total power is divided equally between the two orthogonal waveforms used for the case of 2 bits per pulse while the total power is assigned to a single waveform for the case of 1 bit per pulse. As the data rate increases, the total power is divided amongst more waveforms, causing the respective BER curves to further shift to the right on the SNR axis. The figure also shows that both the 10-antenna sparse array and the 10-antenna ULA yield the same BER performance since the communication direction is well separated from the mainlobe region and the SLLs used for communications are controlled to be the same for both arrays. Although the communications performance is the same, the sparse array configuration has a narrower mainlobe which enables higher angular resolution for the radar operation as compared to the ULA configuration.

\begin{figure}
    \centering
    \begin{minipage}{0.45\textwidth}
        \centering
        \includegraphics[trim = {6cm 10.2cm 7cm 9.8cm}, scale=0.7]{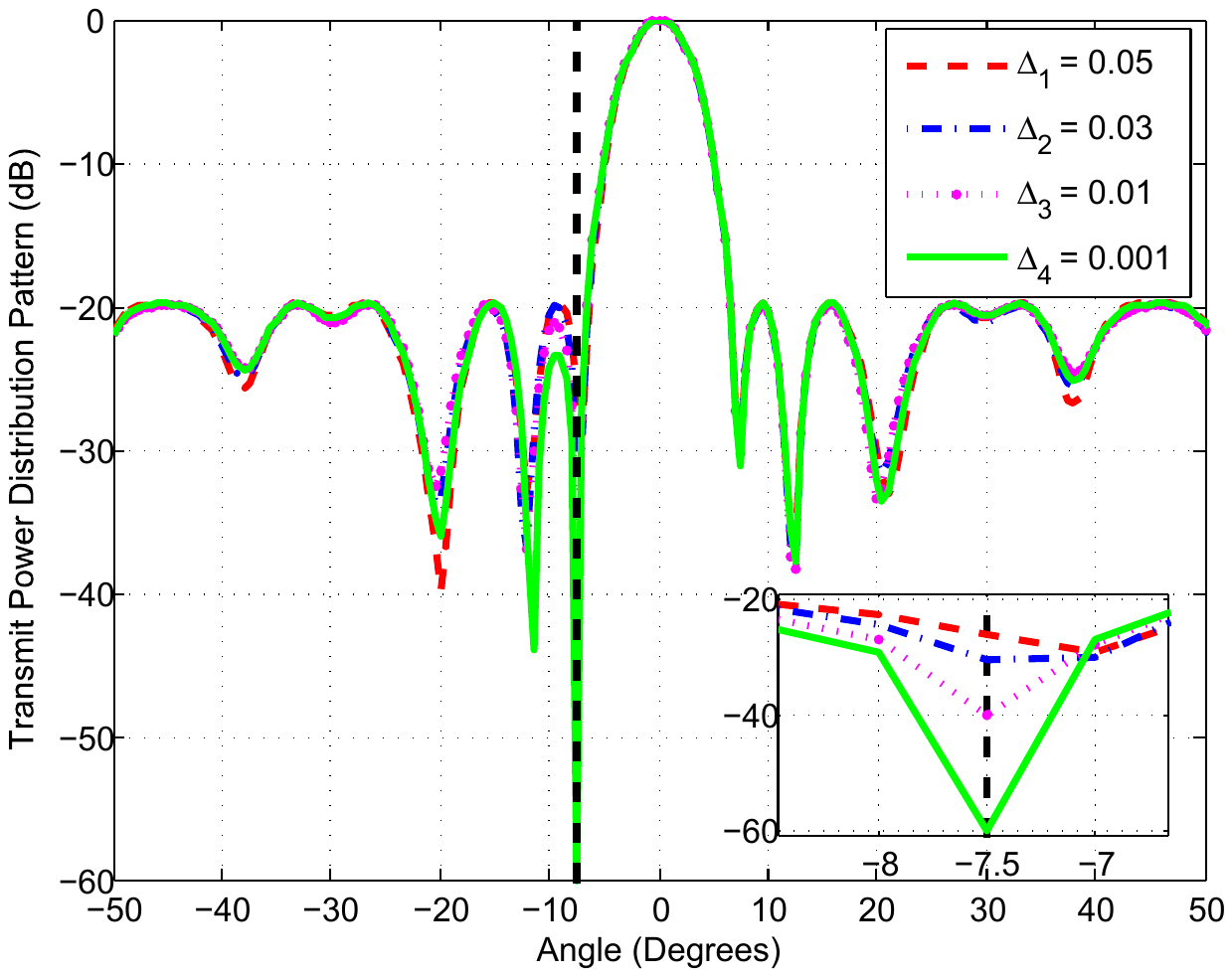}
         \caption{Synthesized focused beampatterns of sparse array (c) for communication symbols $\Delta=\{0.05, 0.03, 0.01, 0.001\}$.}
        \label{fig_5a}
    \end{minipage}\hfill
    \begin{minipage}{0.45\textwidth}
        \centering
        \includegraphics[trim = {3cm 2cm 0cm 0cm}, width=9cm, height=7cm]{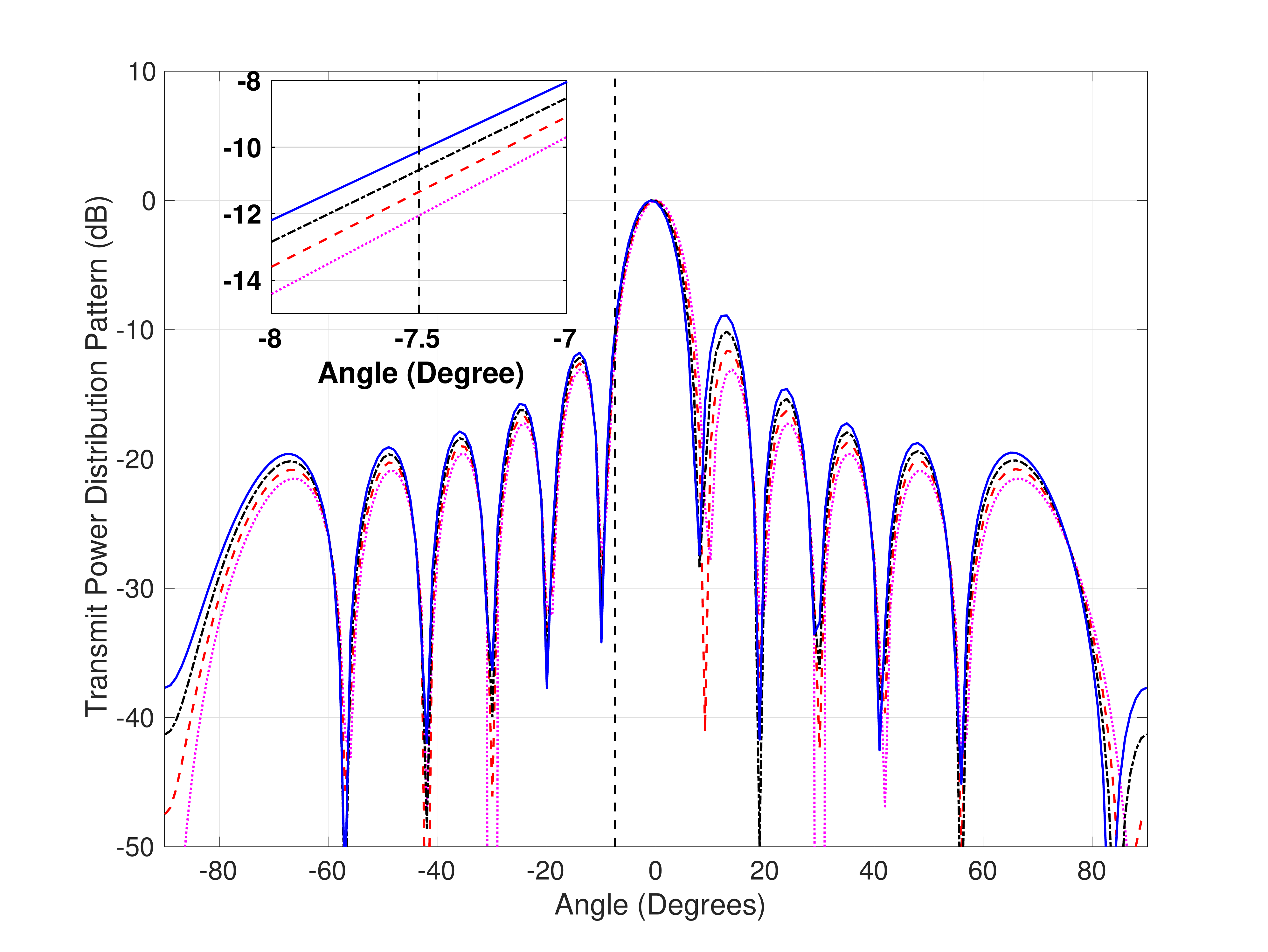} 
        \caption{Synthesized focused beampatterns of 12-antenna ULA for communication symbols $\Delta=\{0.3, 0.28, 0.26, 0.24\}$.}
        \label{fig_5b}
    \end{minipage}
\end{figure}

\begin{figure}
    \centering
    \begin{minipage}{0.45\textwidth}
  		\centering
    	\includegraphics[scale=0.7]{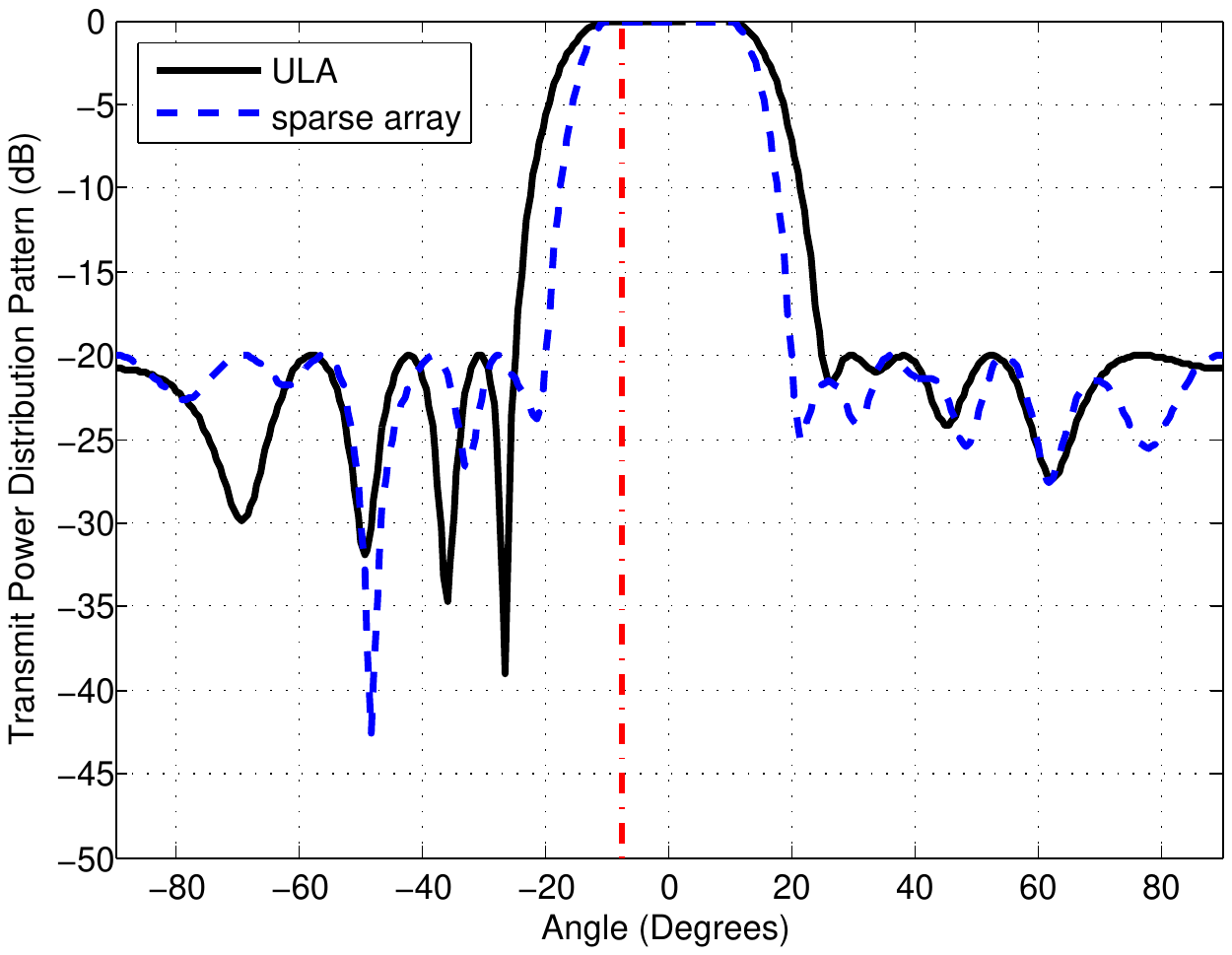}
    	\caption{Synthesized flat-top beampatterns of 12-antenna sparse array (d) and 12-antenna ULA for PM-based signaling scheme.}
\label{fig_5c}
    \end{minipage}\hfill
    \begin{minipage}{0.45\textwidth}
  		\centering
    	\includegraphics[trim = {1cm 1cm 1cm 0.5cm}, scale=0.68]{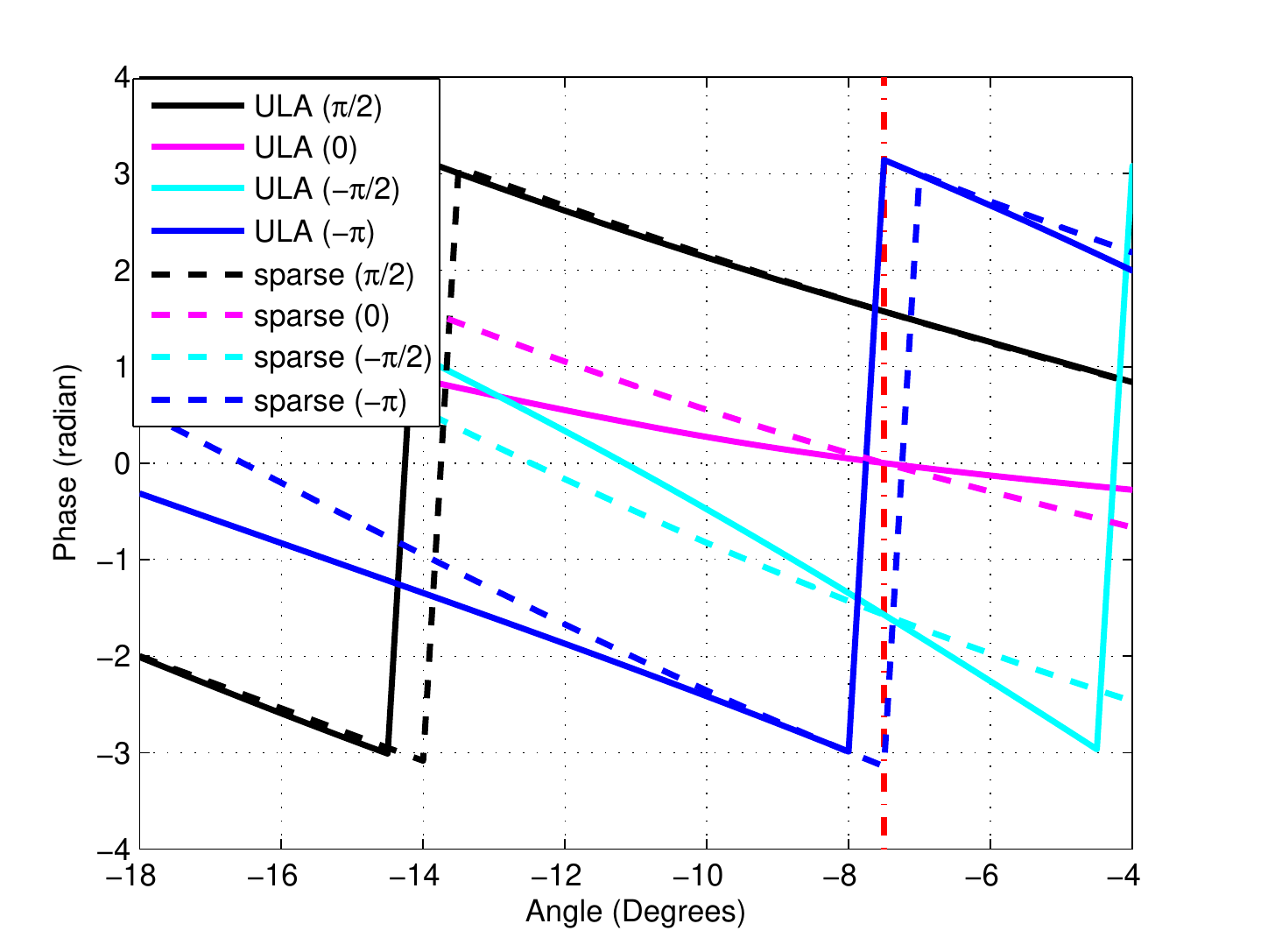}
    	\caption{The phase profiles around the direction of communication receiver for each QPSK symbol $\{-\pi/2, 0, \pi/2, \pi \}$. Note that the phases are wrapped within the range $[-\pi. \pi]$ radian.}
        \label{fig_5d}
    \end{minipage}
\end{figure}

We then change the direction of the communication receiver to be $\theta_c = -7.5^{\circ}$, which is close to the radar target. The number of selected antennas is increased to $12$ and the configurations of two selected 12-antenna sparse arrays are depicted in Figs.~\ref{fig_3}~(c)~and~(d). The sparse array in Fig.~\ref{fig_3}~(c) aims at synthesizing a focused-shape beampattern pointing towards the radar direction $\theta_t=0^\circ$ while ensuring that the communication direction $\theta_c=-7.5^{\circ}$ is sitting in the sidelobe region. We also use for comparison a 12-antenna ULA with inter-element spacing of half wavelength. The beampatterns of the sparse array (c) and ULA with four modulated SLLs are plotted in Fig. \ref{fig_5a} and \ref{fig_5b}, respectively. We can observe that the selected 12-antenna sparse array (c) occupies almost the full array aperture. As a result, the synthesized beampatterns exhibit high angular resolution compared with that of the 12-antenna ULA. Again, the power patterns of the sparse array (c) exhibit almost the same shape with a constant PSL of $-20$dB regardless of different communication symbols, whereas the power level of the ULA reaches $-12.9$dB at the communication angle of $-7.5^{\circ}$. The figures also show that the communication direction is located in the sidelobe region in the case of sparse array (c) while it overlaps with the mainlobe for the ULA. The sparse array (d) is used to synthesize a flat-top beampattern with the communication receiver locating in the mainlobe. Fig. \ref{fig_5c} shows that the power pattern of the sparse array (d) exhibits a sharper transition band compared to that of the ULA, which enables better transmit power concentration and improved robustness. The phase profiles around the direction of the communication receiver for each QPSK symbol are plotted in Fig. \ref{fig_5d}. The figure shows that the phase patterns for  sparse array~(d) change linearly with the arrival angle, which implies constant phase differences between different QPSK symbols. Thus, robust communication performance against the communication receiver angle deviation can be achieved.

%


%
\begin{figure}[!h]
  \centering
    \includegraphics[trim = {6cm 1cm 6cm 0cm}, scale=0.8]{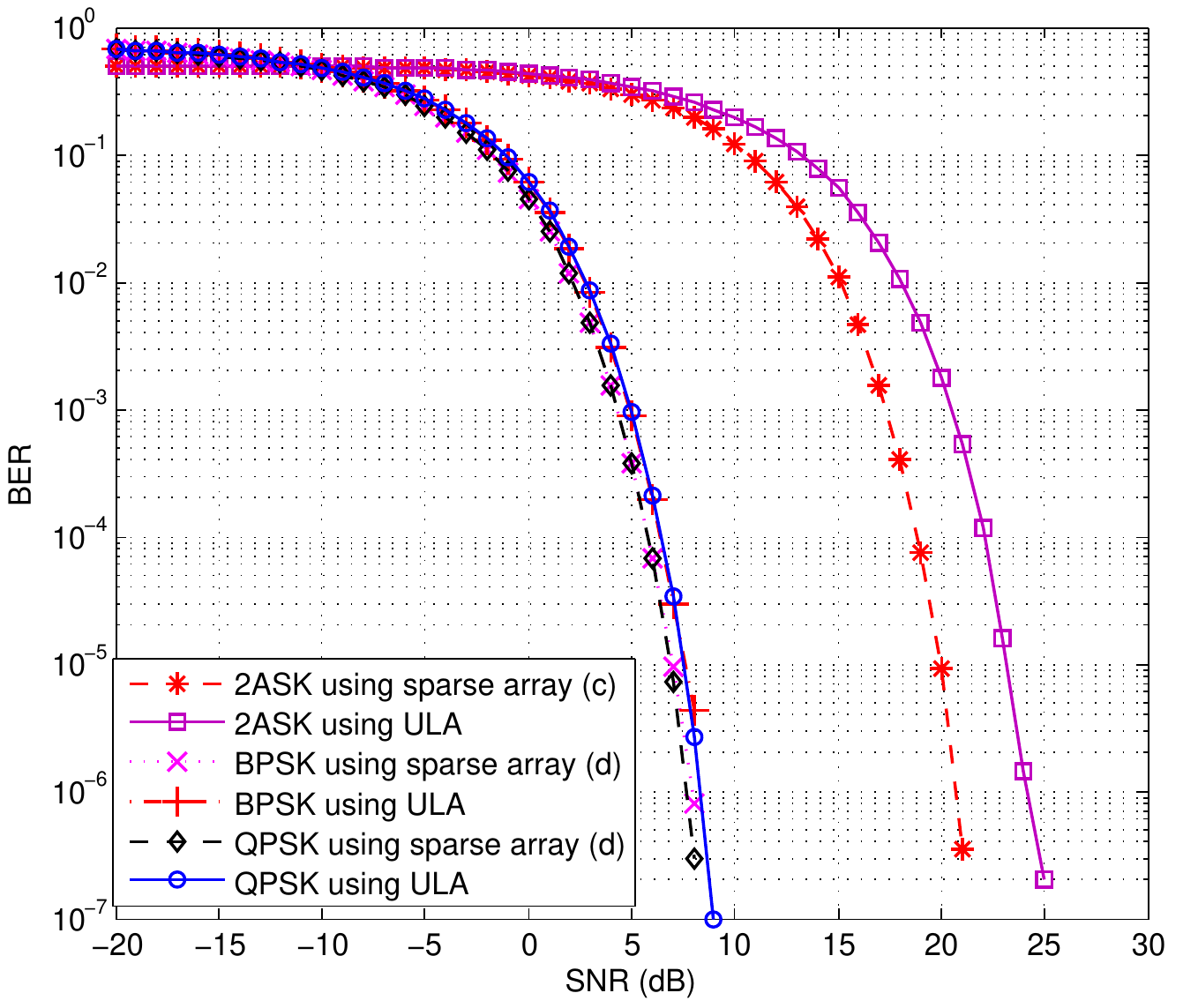}
    \caption{BER versus SNR using 12-antenna sparse arrays (c) and (d) and 12-antenna ULA. Communication receiver locates at direction $\theta_c=-7.5^\circ$.}
\label{BERfig2}
\end{figure}

\begin{figure}[!h]
  \centering
    \includegraphics[trim = {6cm 1cm 6cm 1cm}, scale=0.3]{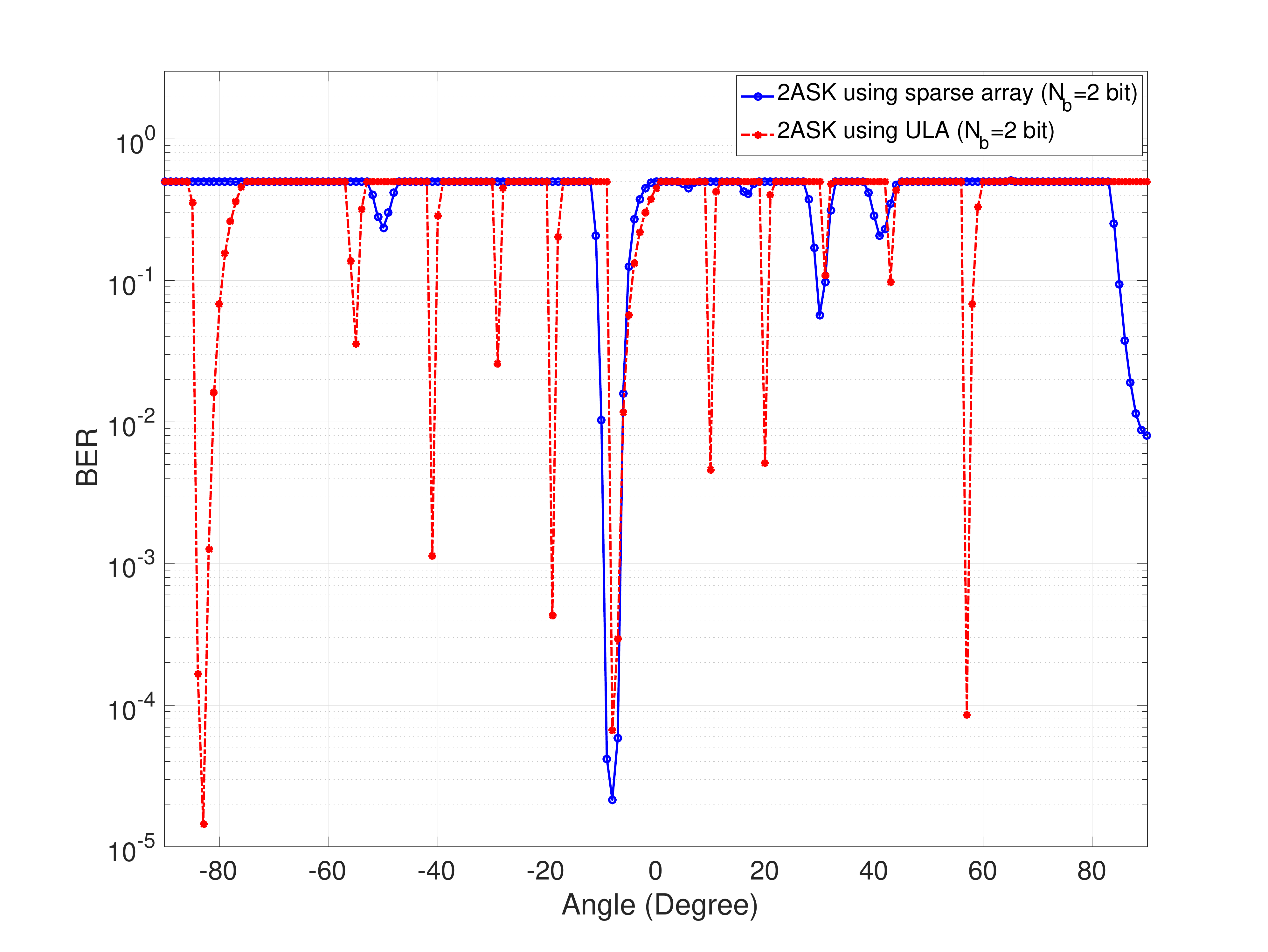}
    \caption{BER versus direction of the communication receiver using 12-antenna sparse array (c) and 12-antenna ULA. SNR=20~dB is used.}
\label{BERfig3}
\end{figure}
Fig. \ref{BERfig2} shows the BER curves versus SNR for information embedding towards a communication receiver located in the direction $\theta_c=-7.5^\circ$. We use multi-waveform 2ASK information embedding for the sparse array and ULA associated with Figs. \ref{fig_5a} and \ref{fig_5b}. Note that the direction $\theta_c=-7.5^\circ$ overlaps with the mainlobe of the ULA while it is well separated from the mainlobe of the sparse array thanks to the narrow mainlobe property of the sparse array. Since the radar operation requires the mainlobe to remain the same during the entire dwell time of the radar operation, the ULA is designed such that the transmit gain associated with binary bit ``0'' equals 0.9 times of the transmit gain associated with binary bit ``1'', i.e., the maximum variation of the transmit gain within the mainlobe of the radar is kept within $10\%$ of its maximum value. A larger variation between the two beampatterns is undesirable and can cause disturbance to the radar operation. Fig. \ref{BERfig2} shows that the BER performance for the sparse array outperforms that associated with the ULA. This can be attributed to the fact that the communication direction is located outside the mainlobe of the sparse array and, therefore, the SLLs used can be well separated from each other. It is worth noting that, if the radar operation requires the mainlobe for the ULA to be exactly same as that of the sparse array, then 2ASK communication technique completely fails.

As the PM signaling scheme enables the concurrent communications within the radar mainlobe direction, we calculate the BER curves versus SNR utilizing the BPSK and QPSK modulations for the sparse array shown in Fig. \ref{fig_3}(d), compared with the 12-antenna ULA. Two orthogonal waveforms and two phase symbols $\{0, -\pi\}$ are used for the BPSK embedding scheme, whereas one orthogonal waveform and four phase symbols $\{-\pi, -\pi/2, 0, \pi/2\}$ are used to test the QPSK signaling scheme. The results are plotted in Fig. \ref{BERfig2}. We can observe that the sparse array exhibits a slightly better communication performance utilizing the two PM signaling schemes. 

The BER is also computed versus angle of the communication receiver using the 2ASK signaling schemes for the sparse array and the ULA associated with the focused transmit beams shown in Figs. \ref{fig_5a} and \ref{fig_5b}. The SNR is fixed to 20 dB. The BER curves for the sparse and the ULA arrays are shown in Fig. \ref{BERfig3}.  The figure demonstrates that the use of sparse array enables communications delivery towards the intended communication direction and does not enable eavesdroppers located at other directions to intercept the embedded data. The figure also shows that the use of ULA allows eavesdroppers located at several directions in the sidelobe region to detect the data. Therefore, the use of sparse arrays for information embedding guarantees more security as compared to ULA.

\begin{figure}[!h]
  \centering
    \includegraphics[scale=0.6]{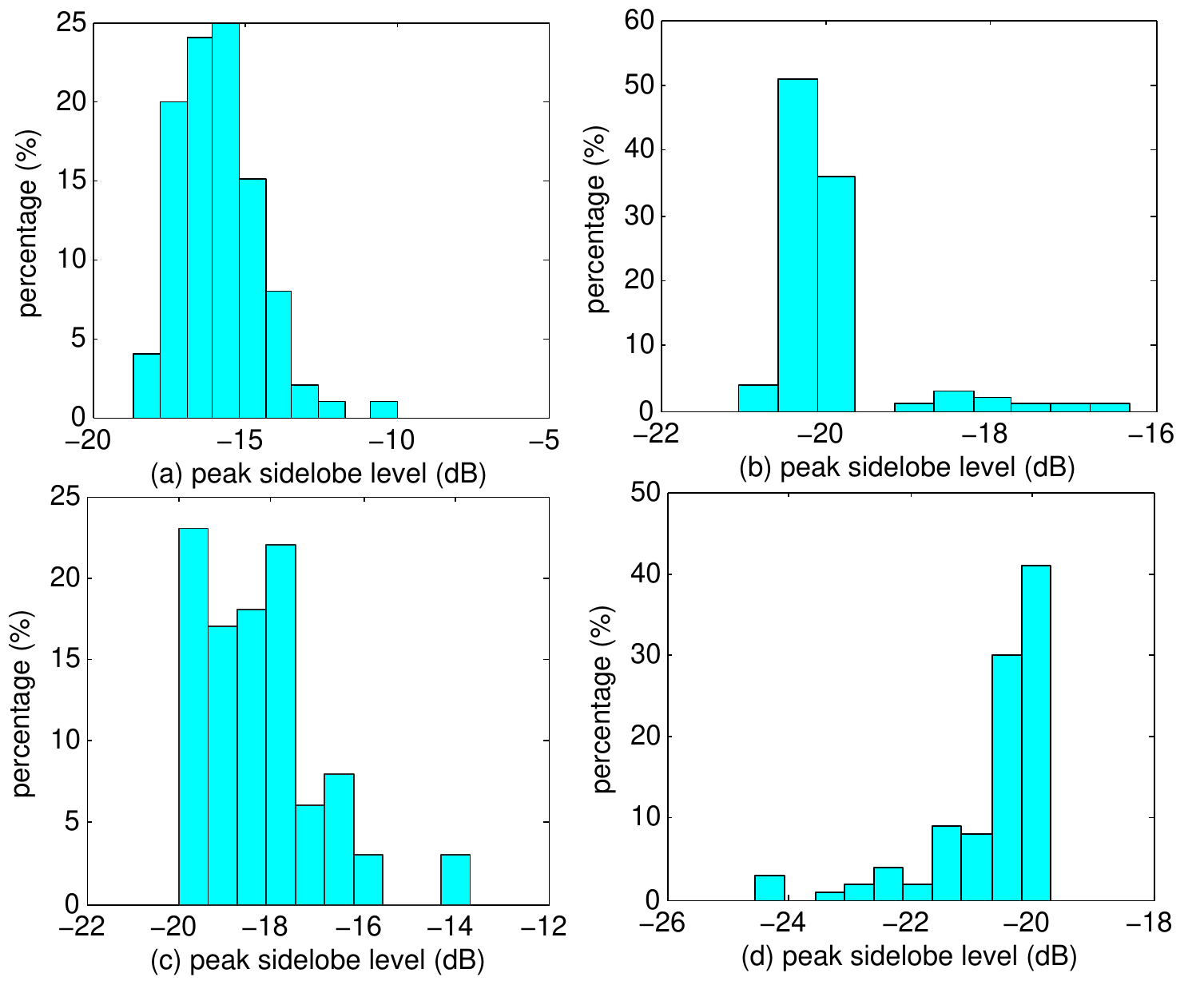}
    \caption{Histograms: (a) PSL of traditional updating rule in the case of focused beam. (b) PSL of proposed updating rule in the case of focused beam. (c) PSL of traditional updating rule in the case of flat-top beam. (d) PSL of proposed updating rule in the case of flat-top beam.}
\label{fig_Hist}
\end{figure}

Finally, we analyze the robustness of the proposed sparse array design method against the choice of initial search points. The communication receiver is assumed to locate at $-7.5^{\circ}$. We compare the sensitivity of the proposed updating rule shown in Eqs. (\ref{eq:expre_rupdate}) and (\ref{eq:expre_rupdate1}) with the traditional method in both cases of focused and flat-top beampattern synthesis. We conduct 100 Monte-Carlo trials with randomly chosen initial search points. The PSL of synthesized beampatterns with different starting points is summarized in the histograms of Fig. \ref{fig_Hist}. We can see that the proposed updating rule can design a sparse array with a satisfied beampattern, regardless of the initial search point, thus exhibiting much better robustness and shorter convergence time compared with the traditional method.

\subsection{Example 2: Common Array Design with Multiple Beamformers}

\begin{figure}[!ht]
  \centering
    \includegraphics[scale=1]{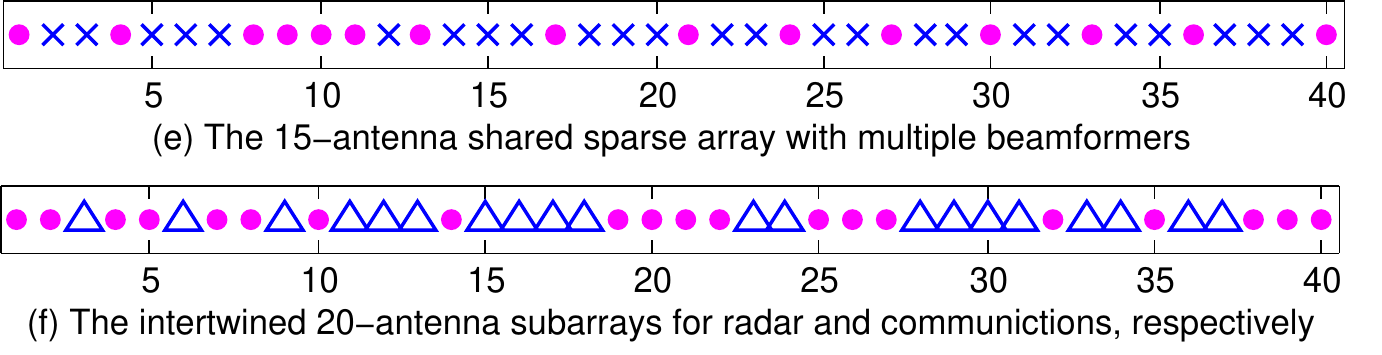}
    \caption{(e) Proposed 15-antenna sparse array for multiple beamformers when $\theta_c=-5^{\circ}$. (f) The two intertwined 20-antenna subarrays for radar and communications, respectively. The subarray indicated by filled circles is for radar function and that indicated by triangles is for communications.}
\label{fig_7}
\end{figure}

We then consider the shared sparse array design with respective beamformers for radar and communications. The radar target and a single communication receiver are assumed at direction $\theta_t=0^{\circ}$ and $\theta_c=-5^{\circ}$, respectively. The sparse array of $15$ antennas is selected and associated with two designed beamformers. One beamformer aims at concentrating the radiation power towards the radar target and forming a complete null towards the communication receiver. The other beamformer strives to minimize the power level within the sidelobe angular sector while maintaining the unit array again towards the communication receiver. As the communication receiver is close to the radar target, a desired beampattern with the half power beam width (HPBW) of around $3^{\circ}$ is required. The structure of the 15-antenna sparse array is depicted in Fig. \ref{fig_7} (e), which spans the full aperture length. The power patterns associated with the two beamformers are depicted in Fig. \ref{fig_81} with a peak SLL of $-15$dB. The 15-antenna ULA with an inter-element spacing of half wavelength fails to synthesize a narrow mainlobe towards the target while simultaneously forming a deep null in the communication direction due to its low spatial resolution. We then plot the power patterns of a 20-antenna ULA with inter-element spacing of $\lambda/2$ in Fig. \ref{fig_82} for comparison. The source efficiency of deploying sparse arrays for DFRC systems are clearly manifested from the comparable performance of the 20-antenna ULA with that of the 15-antenna sparse array. It is worth noting that plotting BER curves becomes unnecessary for performance validation here, as communication symbols are no longer needed to embed into the radar pulse transmitting due to independent beamformers and waveform diversity.

\subsection{Example 3: Intertwined Subarray Design with Shared Aperture}

\begin{figure}
    \centering
    \begin{minipage}{0.45\textwidth}
        \centering
        \includegraphics[scale=0.6]{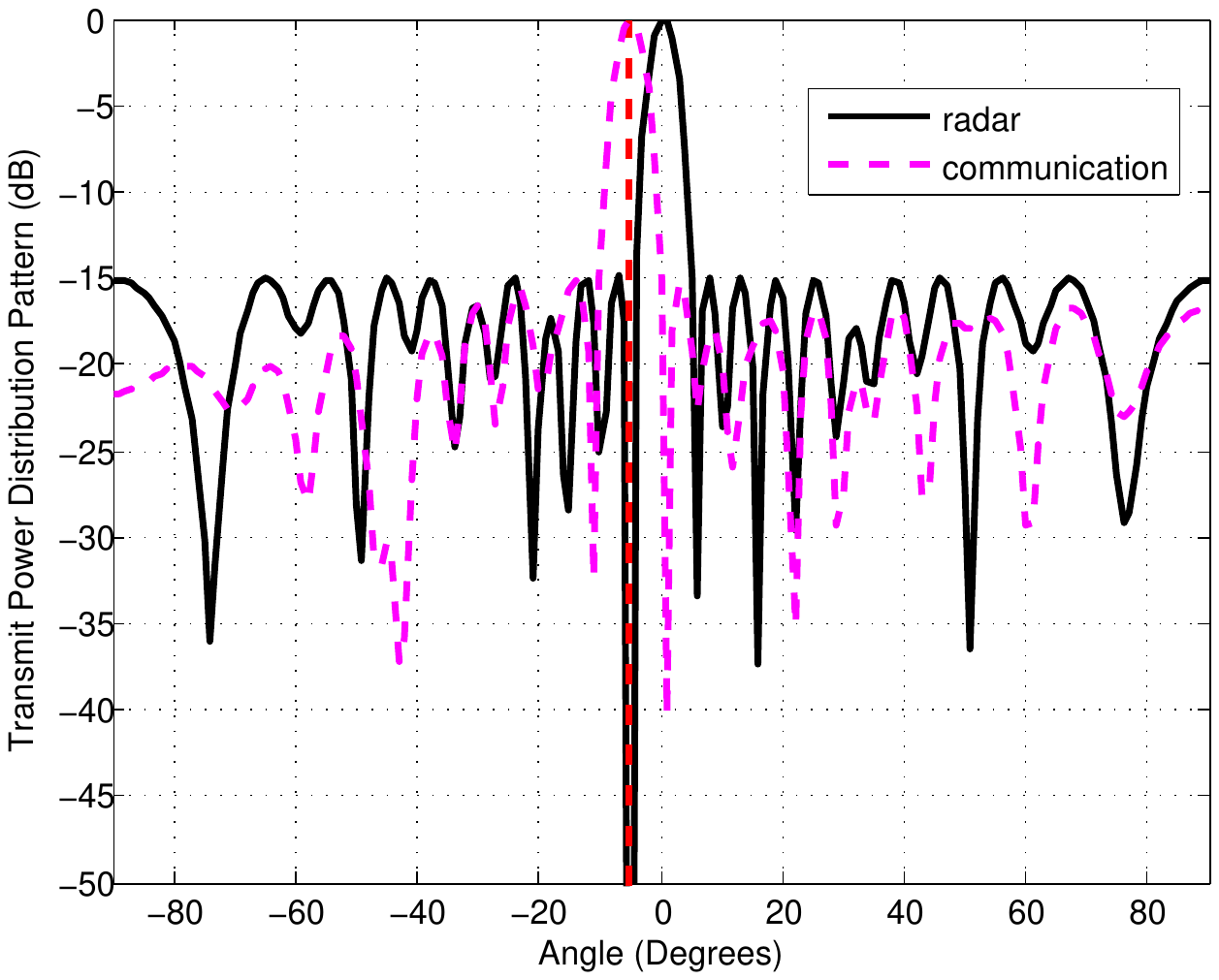} 
        \caption{Power patterns of common 15-antenna sparse array associated with multiple beamformers.}
        \label{fig_81}
    \end{minipage}\hfill
    \begin{minipage}{0.45\textwidth}
        \centering
        \includegraphics[scale=0.6]{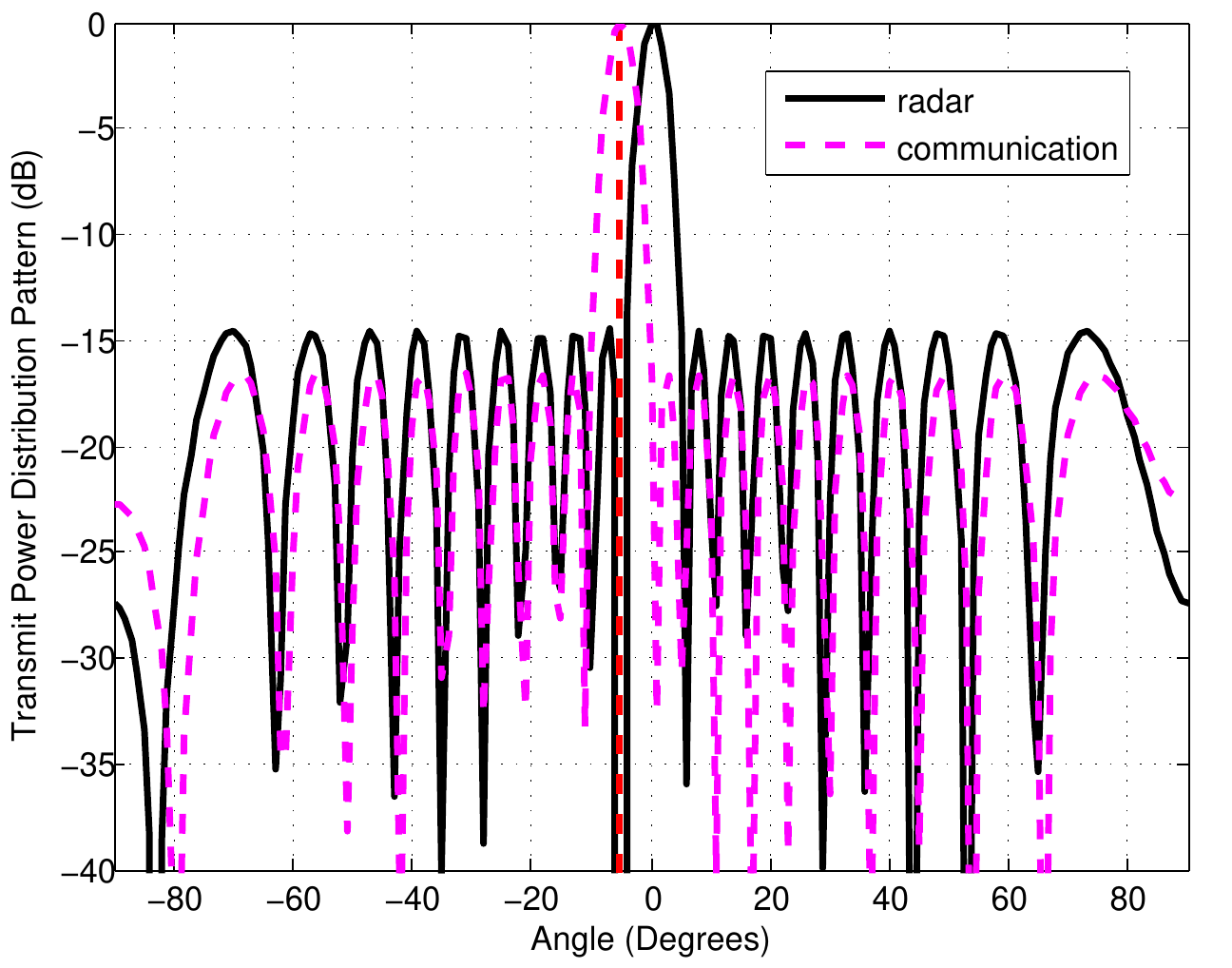} 
        \caption{Power patterns of common 20-antenna ULA associated with multiple beamformers.}
        \label{fig_82}
    \end{minipage}
    \begin{minipage}{0.45\textwidth}
        \centering
        \includegraphics[scale=0.6]{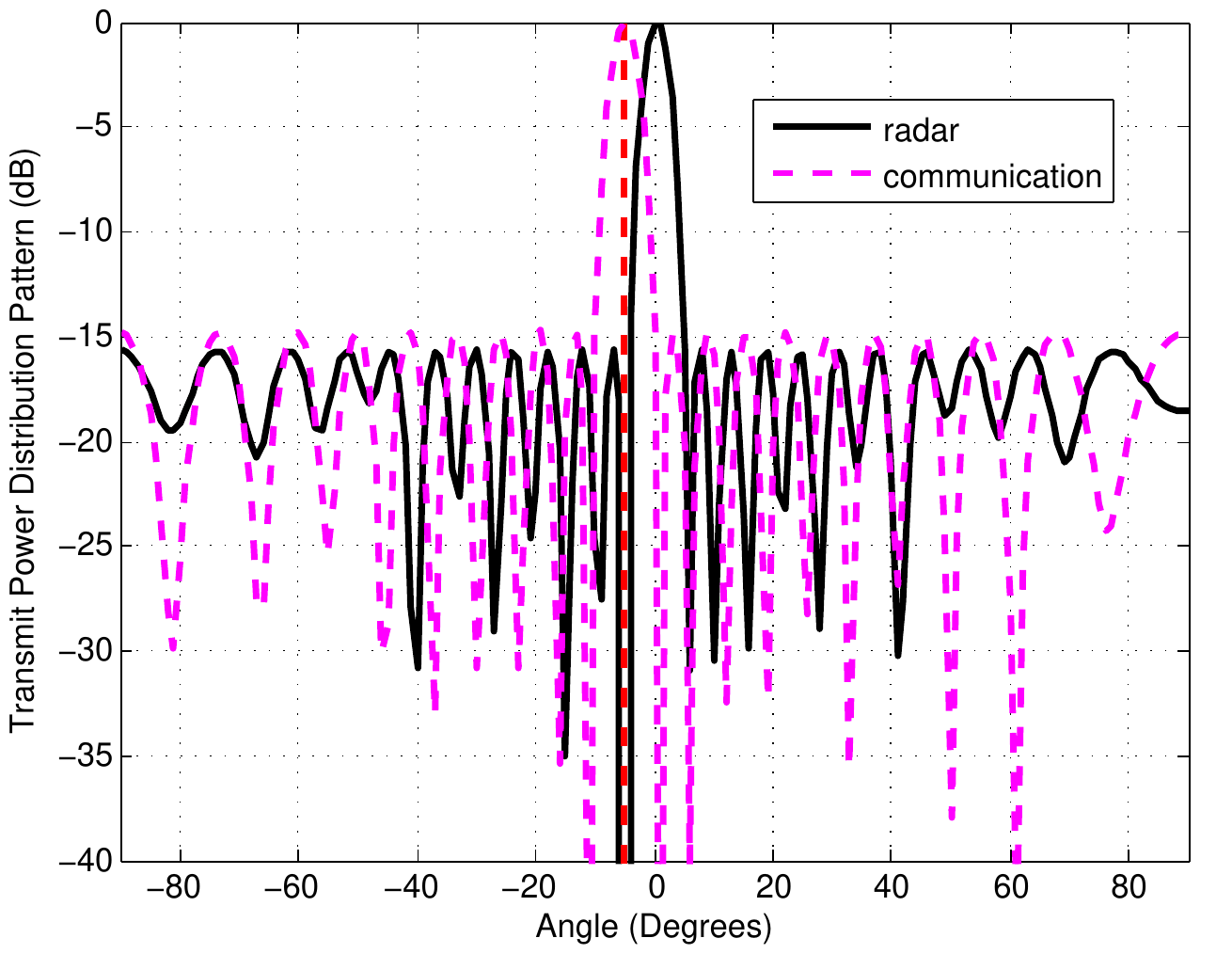} 
        \caption{Power patterns of intertwined 20-antenna subarrays for radar and communications.}
        \label{fig_91}
    \end{minipage}\hfill
    \begin{minipage}{0.45\textwidth}
        \centering
        \includegraphics[scale=0.6]{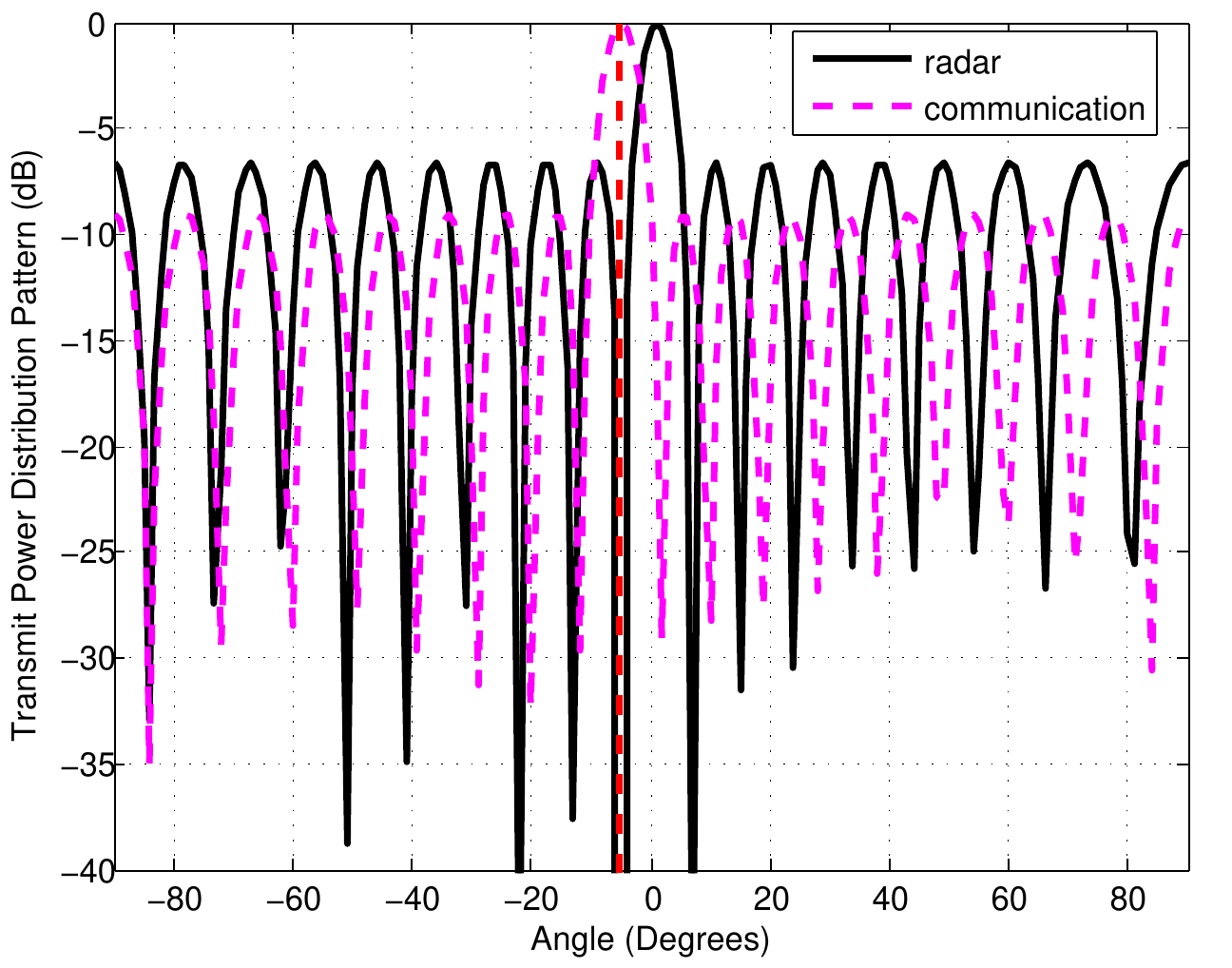} 
        \caption{Power patterns of split 20-antenna subarrays for radar and communications.}
         \label{fig_92}
    \end{minipage}
\end{figure}

In this example, we proceed to investigate the intertwined sparse array design with shared aperture for radar and communication functions. The available $40$ antennas are partitioned into two sparse subarrays: one for radar function and the other for communications. The optimum subarrays will most likely be entwined and not necessarily separated, or split. The two subarrays are plotted in Fig. \ref{fig_7} (f) and their corresponding beampatterns are depicted in Fig. \ref{fig_91}. We can observe that the subarray for radar function forms a narrow beam towards the target and a deep null towards the communication receiver. The subarray for communications remains a constant $-15$dB power level in the sidelobe angular region including the radar target direction. We also plot the power patterns of two split subarrays in Fig. \ref{fig_92} for comparison, where the first 20 antennas compose a subarray for radar function and the remaining 20 antennas compose the other subarray for communications. We can observe that the radiation patterns of two split subarrays exhibit high SLLs and wide mainlobes due to its low spatial resolution caused by limited aperture length, which will undoubtedly affect the normal radar function.

We calculate the aperture efficiency of sparse arrays (a), (c), (e), (f) and associated weights in Table \ref{table_2}. The aperture efficiency is defined as the ratio between the directivity of sparse array and that of uniformly excited ULA \cite{Hansen2009}, which determines the directivity loss due to the non-uniform tapering. It is defined as $\eta = \frac{G_{\rm s}}{G_0}$, with
\begin{equation}
\label{eq:expre_uniform}
G_0 =\frac{M^2}{M+2\sum_{m=1}^{M-1} (M-m) \rm sinc (mk_0d) },
\end{equation}
and
\begin{equation}
\label{eq:expre_taper}
G_{\rm s} = \frac{|\sum_m^M w_m|^2}{\sum_{n=1}^M \sum_{m=1}^M w_nw_m^* \rm sinc((n-m)k_0d)}.
\end{equation}
We can see from Table \ref{table_2} that the aperture efficiency of proposed sparse arrays are relatively high.
\begin{table}[!h]
\caption{Aperture efficiency of sparse arrays (a), (c), (e), (f).}
\label{table_2}
\begin{center}
\begin{tabular}{|c|c|c|c|}
\hline
\hline
array (a) & array (c) & array (e) & array (f)   \\
\hline
0.95 & 0.91 & 0.86 & 0.87\\
\hline
\hline
\end{tabular}
\end{center}
\end{table}

\section{Conclusions}
\label{sec:conclusions}

In this paper, we addressed the problem of sparse array design by antenna selection under the framework of dual functional radar communications systems. The additional spatial DoFs and configuration flexibility provided by sparse arrays were utilized to suppress the cross-interference and facilitate the cohabitation of the two functions. We solved the new problem of integrated sparse array design and transmit beampattern synthesis with additional constraints imposed by the co-design of two simultaneous functions. To increase the robustness of antenna selection algorithm against initial search points, we proposed a new selection vector updating rule. Furthermore, we proposed two new dual-function systems with a common sparse array associated with different beamformers and complementary sparse arrays with shared aperture. Simulation results showed that sparse arrays can synthesize a narrow mainlobe with well-controlled sidelobes, thus increasing the spatial resolution and suppressing cross-interference between the two simultaneous functions. Moreover, the comparable performance exhibited by fewer-antenna sparse arrays with large ULAs demonstrated their advantages in resource management and hardware efficiency.

\bibliographystyle{ieeetr}
\bibliography{biblio}

\end{document}